\documentclass[useAMS,usenatbib]{mn2e}
\usepackage{txfonts}
\usepackage{graphicx}
\usepackage{multirow}
\usepackage{dcolumn}
\usepackage{array}
\usepackage{xcolor}
\usepackage{float}

\title[The formation mechanism of 4179 Toutatis]{The formation mechanism of 4179 Toutatis' elongated bi-lobed structure in a close Earth encounter scenario}

\author[S.C. Hu et al.]{
Shoucun Hu,$^{1, 2}$
Jianghui Ji,$^{1}$\thanks{E-mail: jijh@pmo.ac.cn}
Derek C. Richardson,$^{3}$
Yuhui Zhao,$^{1}$
Yun Zhang$^{4}$
\newauthor{
}
\\
$^{1}$CAS Key Laboratory of Planetary Sciences, Purple Mountain Observatory, Chinese Academy of Sciences, Nanjing 210008, China\\
$^{2}$University of Chinese Academy of Sciences, Beijing 100049, China\\
$^{3}$Department of Astronomy, University of Maryland, College Park, MD 20740-2421, United States\\
 $^{4}$School of Aerospace Engineering, Tsinghua University, Beijing 100084, China\\
}
\date{Received 2014 December 14; in original form 2014 December 30}

\begin{document}
\label{firstpage}
\pagerange{\pageref{firstpage}--\pageref{lastpage}}\pubyear{2015}
\maketitle

% Abstract of the paper
\begin{abstract}
The optical images of near-Earth asteroid 4179 Toutatis acquired by Chang'e-2 spacecraft show that Toutatis has an elongated contact binary configuration, with the contact point located along the long axis. We speculate that such configuration may have resulted from a low-speed impact between two components. In this work, we performed a series of numerical simulations and compared the results with the optical images, to examine the mechanism and better understand the formation of Toutatis. Herein we propose an scenario that an assumed separated binary precursor could undergo a close encounter with Earth, leading to an impact between the primary and secondary, and the elongation is caused by Earth's tide. The precursor is assumed to be a doubly synchronous binary with a semi-major axis of 4 $R_p$ (radius of primary) and the two components are represented as spherical cohesionless self-gravitating granular aggregates. The mutual orbits are simulated in a Monte Carlo routine to provide appropriate parameters for our \emph{N}-body simulations of impact and tidal distortion. We employ the \emph{pkdgrav} package with a soft-sphere discrete element method (SSDEM) to explore the entire scenarios. The results show that contact binary configurations are natural outcomes under this scenario, whereas the shape of the primary is almost not affected by the impact of the secondary. However, our simulations further provide an elongated contact binary configuration best-matching to the shape of Toutatis at an approaching distance $r_p$ = 1.4 $\sim$ 1.5 $R_e$ (Earth radius), indicative of a likely formation scenario for configurations of Toutatis-like elongated contact binaries.
\end{abstract}

% Select between one and six entries from the list of approved keywords.
% Don't make up new ones.

%%%%%%%%%%%%%%%%%%%%%%%%%%%%%%%%%%%%%%%%%%%%%%%%%%

\section{Introduction}
Asteroids are remnant building blocks of the formation of our Solar System, and provide key clues to planet formation and the origin of life on the Earth. Among them, near-Earth asteroids (NEAs) are of particular concern, since they approach Earth at a relatively close distance or even impact the Earth.

4179 Toutatis is an S-type NEA with a highly eccentric orbit, approaching Earth every four years between 1992 and 2012, and currently is in a 1:4 mean-motion resonance with Earth \citep{whipple1993long, michel1996dynamical}. Toutatis has become the focus of several ground-based observational campaigns using different techniques (e.g., photometry, spectroscopy, and radar) since it was discovered in 1989. Spectral analysis suggests that the surface composition of Toutatis is consistent with an L chondrite type assemblage, with an estimated bulk density of $\sim 2.1-2.5$ g cm$^{-3}$ \citep{reddy2012composition}. Numerical investigations of long-term orbit evolution show that Toutatis' orbit is very chaotic and the Lyapunov time is only about 50 years \citep{whipple1993long, michel1996dynamical}. This is because the very low inclination of the asteroid (only $0.47^\circ$) can lead to frequent close approaches with terrestrial planets. The possibility of Toutatis-Earth impact is also not excluded but it is highly unpredictable \citep{sitarski1998motion}. According to the origin scenario of NEAs, Toutatis may come from the main-belt and was transported to NEA space by a 3:1 orbital resonance with Jupiter \citep{whipple1993long, krivova1994orbit}, secular resonances \citep{michel1996dynamical,Ji2001} and further driven by Yarkovsky effect \citep{bottke2002debiased, morbidelli2003yarkovsky, bottke2006yarkovsky}.

From the lightcurves of Toutatis, \cite{spencer1995lightcurve} estimated that this asteroid has a complex rotation with a period between 3 and 7.3 days. Utilizing radar observations obtained from Arecibo and Goldstone, \cite{hudson1995shape} and \cite{ostro1995radar} reconstructed the three-dimensional shape and spin state of Toutatis. The shape model indicates that Toutatis has a contact binary-like configuration with dimensions along the principal axes of (1.92, 2.40, 4.60) $\pm$ 0.10 km. They also showed that Toutatis is a non-principal-axis rotator: it rotates along its long axis with a period of 5.41 days (${P_\psi }$) and the long axis precesses with a period of 7.35 days (${P_\phi }$) around the angular momentum vector. \cite{scheeres2000effects} found that a single and strong interaction with a planet can greatly change an asteroid's spin state, causing it to tumble and significantly increase or decrease its angular momentum, which may be used to interpret the current spin state of Toutatis. Combined the data from \cite{spencer1995lightcurve} and \cite{hudson1995shape}, \cite{hudson1998photometric} refined the spin state slightly. Their analysis indicated that a fine regolith layer covers a large part of the surface of Toutatis. Incorporated with the radar observations acquired during the 1996 near-Earth approach,  \cite{ostro1999asteroid} obtained refined estimates of the asteroid's orbit, spin state, and surface properties. In their work, the two periods are refined as ${P_\psi } = 5.367$ days and ${P_\phi } = 7.420$ days by combining the optical and radar data. In addition, results showed that Toutatis has strikingly uniform surface characteristics at centimeter-to-decameter scale, which provides evidence for the presence of regolith. As shown in the panel (a) of Fig. \ref{fig:shape}, a more precise model with resolution $\sim$ 34 m was given by \cite{hudson2003high}, in which concavities, linear and curvilinear ridges, and grooves are shown, implying a possibly complex formation history of Toutatis.

On 13 December 2012, the first close observation of Toutatis was accomplished by Chang'e-2 spacecraft, at a flyby distance of 770 $\pm$ 120 m from the asteroid's surface and a high relative velocity magnitude of 10.73 km/s \citep{huang2013ginger,huang2013engineering, zou2014preliminary,Ji2016}. More than 300 optical images with spatial resolutions from 2.25 m to 80 m per pixel were acquired by utilizing a solar-wing panel's monitoring camera during the outbound flyby; about 45\% of Toutatis' surface was imaged \citep{huang2013ginger,huang2013engineering}. The first panoramic image was taken at a distance of 67.7 km with a resolution of 8.30 m, which is shown in the panel (b) of Fig. \ref{fig:shape}. The maximum physical length and width are then estimated to be $4.75 \times 1.95$ km $\pm$ 10\% \citep{huang2013ginger}. The bifurcated configuration composed of two lobes (body and head) can be seen in the images than in the radar model, implying a contact binary for Toutatis \citep{huang2013ginger, zhu2014morphology, Ji2016}. A giant depression with diameter 800 m at the big end, a sharply perpendicular silhouette near the neck region, and obvious evidence of boulders and craters are evident in the image, which implies that Toutatis may occupy a rubble-pile structure \citep{huang2013ginger, zhu2014morphology, Ji2016}. \cite{zhu2014morphology} proposed that the giant depression may be a crater, which was formed by an impactor with a diameter of $\sim$ 50 m traveling at 5 km/s  by applying a strength-scaling relationship \citep{holsapple2007crater}. Using the optical images of Toutatis, combined with the direction of the camera's optical axis and the relative flyby velocity, the orientation of long axis and the conversion matrix from the inertial system to the body-fixed frame were calculated \citep{zou2014preliminary, bu2014new,zhao2015orientation,Hu2017}. Moreover, \cite{zhao2015orientation} utilized a least-squares method to determine the rotational parameters and spin state of Toutatis, which confirmed earlier results \citep{ostro1999asteroid, takahashi2013spin}. Using the images, \cite{jiang2015boulders} identified over 200 boulders and plotted their cumulative  size-frequency distribution (SFD). A steep slope was noted in the SFD plot, which indicates a high degree of fragmentation of Toutatis \citep{Barucci2015,Ji2016}. Similar to Itokawa, it is speculated that most boulders on Toutatis' surface may not solely be explained by impact cratering from a scaling law prediction \citep{jiang2015boulders,Ji2016}. Recently, using a new data fusion method in frequency domain, \cite{zhao2016radar} constructed a new surface model of Toutatis by integrating radar data and the images, so that more detailed surface characteristics are reflected in the model.

Radar observations indicate that at least 10\% of NEAs larger than 200 m in diameter are contact binaries \citep{benner2006near,Benner2015}, which places important constraints on their origin and evolution. It is speculated that low-speed impact is regarded as the likely mechanisms to generate this kind of asteroid (or comet) \citep{harmon2010radar, brozovic2010radar, magri2011radar, busch2012radar}. Moreover, close images of 67P/Churyumov-Gerasimenko (67P) by Rosetta showed that the cometary nuclei of 67P bears an obvious bi-lobed structure \citep{jutzi2015shape}. Recently, \cite{jutzi2017formation} investigated 67P's structure by using a smoothed particle hydrodynamics (SPH) shock physics code, which showed a possible formation process of the nuclei resulting from a sub-catastrophic impact. \cite{schwartz2018catastrophic} further combined the \textit{pkdgrav} and SPH code and investigated a possibility that the bi-lobed structure of 67P (and other bi-lobed or elongated comets) can be formed after a catastrophic collision between large bodies while maintaining its volatiles and low density during the process, which implies that the observed prominent geological features on 67P may not be primordial.

In this study, we mainly focus on the formation of the bi-lobe configuration of Toutatis. The absence of large variations of surface color in the images suggests that the body and head of Toutatis may have similar surface composition and thus the same precursor body \citep{huang2013ginger, zhu2014morphology}. The speculated formation scenarios of Toutatis invoking low-speed impact have been described in several previous studies \citep{huang2013ginger, zhu2014morphology}. However, considering the slow spin rate, the contact point located on the high terrain of the body suggests that this configuration is not in a most stable state \citep{scheeres2007rotational}. Accordingly, it is very interesting to carry out extensive numerical simulations to investigate the entire formation process of the bi-lobe geological features.

Tidal effects of terrestrial planets play an important role in the evolution of NEAs. For very close encounter distances, this mechanism may shatter, elongate or resurface a NEA \citep{richardson1998tidal, walsh2006binary, walsh2008steady, yu2014numerical} and eventually lead to an escape or a mutual impact between primary and secondary for a binary \citep{chauvineau1991lifetime, chauvineua1995evolution, farinella1992evolution, farinella1993evolution, fang2011binary}. \cite{fang2011binary} studied the effect and frequency of close terrestrial planetary encounters and found that close approaches (less than 10 Earth radii) with Earth occur for almost all binary asteroids on the timescales of 1-10 Myr. In the present study, we assume that the precursor of Toutatis is a separated binary system, which approached Earth at a close distance so that two components impacted mutually at a low speed. Assuming each component of the binary has a cohesionless self-gravitational rubble-pile structure, the soft-sphere discrete element method (SSDEM) is applied to simulate the evolution process, including the scenarios of impact and potential tidal distortion. A Monte Carlo routine is used herein to reduce the dimension of the parameter space in our numerical simulations. Based on the simulations of various initial parameters, we aim to shed light on the formation of Toutatis from low-speed impact between two components when making close approaches to Earth. Furthermore, our investigations can further improve understanding of the role of Earth's tides over the evolution of the population of near-Earth binaries, thereby providing clues to formation of contact binaries.

The structure of this work is as follows. Section 2 elaborates on the model and method we employ in this work, and introduces the physical parameters of our assumed Toutatis precursor, SSDEM model, initial parameter space, and dynamical model adopted in the Monte Carlo simulations. Section 3 provides our \emph{N}-body simulation results, which are grouped into 2 categories according to the strength of tidal force. The obtained configurations of newly formed contact binaries are analyzed and compared with optical images from Chang'e-2 spacecraft. The alteration of spin state during encounter is also discussed, which is helpful for understanding the outcomes of low-speed impact in this scenario and possibly give insight into the formation of Toutatis. Finally, the discussion and conclusion are summarized in Section 4.

\section{Model and method}
\subsection{Shape and rotation}
\begin{figure*}
\includegraphics[width=0.80\textwidth]{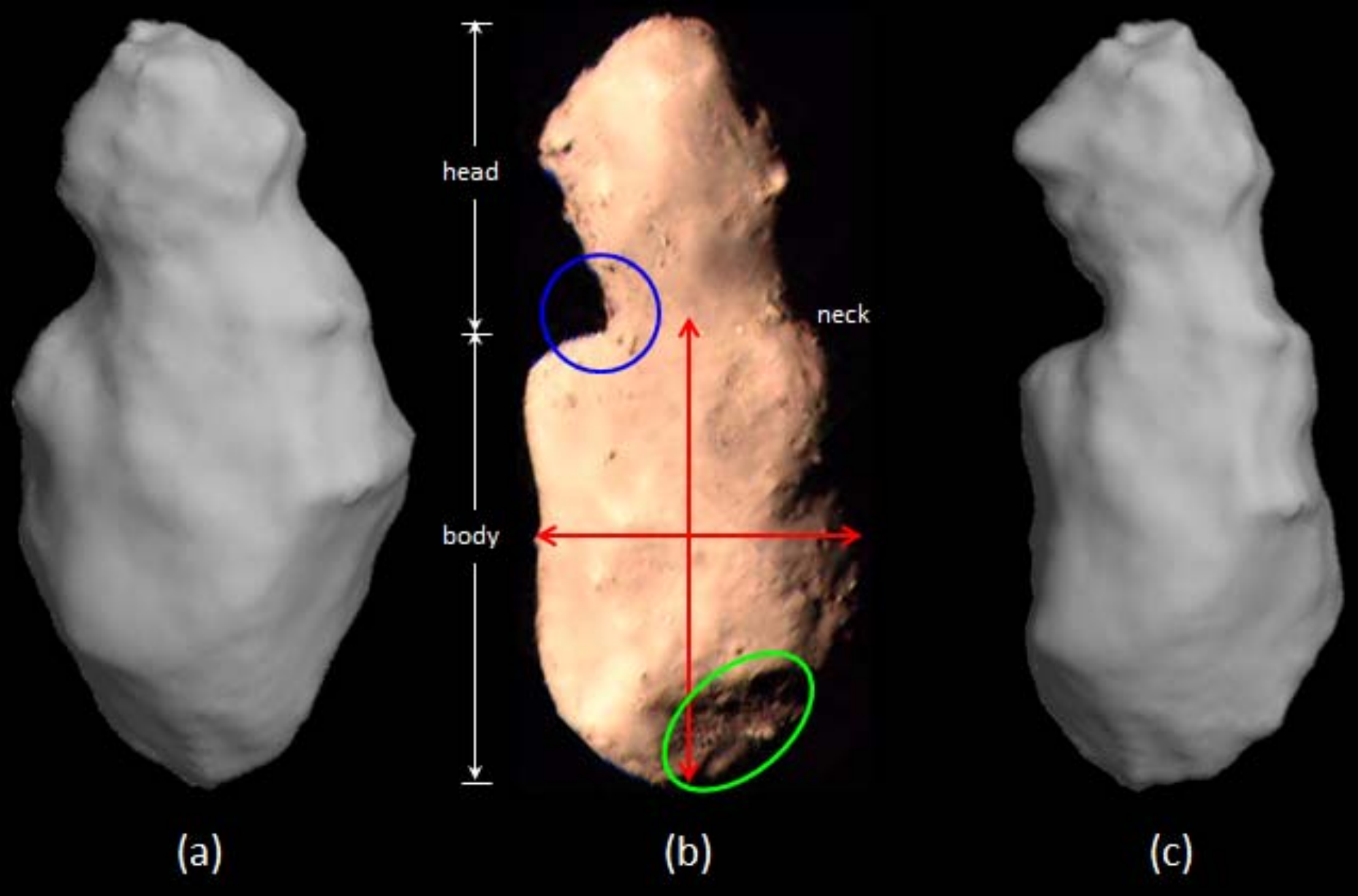}
\caption{The shape of 4179 Toutatis. They are (a): radar model with 12,796 vertexes and 6,400 facets (Hudson et al. 2003); (b): optical image acquired by Chang'e-2; (c): combined model using the optical image and radar model. A perpendicular angle near the neck (blue circle), elongated body (red arrow lines) and a depression near the big end (green ellipse) are marked on the optical image. The radar model has been rotated to match the images.}
\label{fig:shape}
\end{figure*}

The radar model derived by \cite{hudson2003high} is shown in the panel (a) of Fig. \ref{fig:shape}, and the model is rotated to match the attitude to the case shown in the optical image from Chang'e-2 in the panel (b). The distinct perpendicular angle in the neck terrain, the giant depression in the big end, and the elongated body and head are indicated in the optical image. The visual elongation ratio of the body is about 1.5, and the contact point locates along the long axis of the body. As aforementioned, we aim to examine the scenario of a low-speed impact for formation of Toutatis, following a close Earth encounter of an assumed binary precursor. In our investigation, the entire encounter process is simulated using a SSDEM code. However, it is not practical for us to reproduce all configuration characteristics of the images in our simulations. The similarities between our results and the optical image will be compared visually, in which a distinct contact binary configuration, a body shape with elongation ratio $\sim 1.5$ and a contact point along the long axis of the body are the three major characteristics, which we will adopt them to evaluate the results.

In order to match the optical observations as close as possible, we have reconstructed the radar model slightly to match the optical image. This is done by adjusting the $x$ and $y$ coordinates of vertices in the shape model to fit the silhouette of the optical image (in this work, we have ignored the possibility that some portion of Toutatis may not be visible due to shadowing effects). The modified 3D model ("combined model") is shown in the panel (c) of Fig. \ref{fig:shape}. Using the combined model, the mass ratio of the head to the body ($q$) is estimated to be $\sim$0.28 by assuming equal bulk density for two lobes (2.1 g cm$^{-3}$). The primary and secondary precursor are modeled as spherical bodies and their radii ($R_p$ and $R_s$) are calculated so that their volumes are equal to that of the combined model, which are estimated to be 1.09 km and 0.71 km, respectively. The escape speed of the binary at $r = R_p + R_s$ is 1.0 m/s.

In this study, we do not take Toutatis' current peculiar tumbling rotation as a constraint in our simulations, because the rotation can be easy to be altered by the mechanisms such as the YORP effect, or terrestrial planet approaching at a relatively large distance does not change the physical configuration. Nevertheless, we calculate the variations of spin state over the evolution to better understand the outcomes of low-speed impacts in our scenario, giving implication for formation of Toutatis.

\subsection{SSDEM and \emph{pkdgrav}}
In this work, we carry out extensive numerical simulations by using \emph{pkdgrav} package, which is a parallel \emph{N}-body gravity tree code that was designed initially to perform cosmological simulations \citep{stadel2001cosmological}. \cite{richardson2000direct, richardson2009numerical, richardson2011numerical} utilized the code for particle collisions study, in which collisions were treated as instantaneous single-point-of-contact impacts between rigid spheres. The motion of each particle was governed by gravitational forces and contact forces imposed by other particles. A soft-sphere discrete element method (SSDEM) was added by \cite{schwartz2012implementation}, which was tested by reproducing the dynamics of flows in a cylindrical hopper. In this implementation, contacts can last many time-steps and the reaction force is related to the degree of overlap and contact history between particles. A spring-dashpot model is employed --- described in detail by \cite{schwartz2012implementation}, in which contact forces are functions of normal and tangential coefficients of restitution ($\varepsilon_n$, $\varepsilon_t$), normal and tangential stiffness parameters ($k_n$, $k_t$), the static, rolling, and twisting friction coefficients ($\mu_s$, $\mu_r$, $\mu_t$), respectively. Herein we employ an updated SSDEM model, in which a shape parameter $\beta$ is involved to represent the statistical effect of real particle shape \citep{Da2005rheophysics, Estrada2011identification, zhang2017creep}.

We assume that each component of the precursor binary has a rubble pile structure. The primary and secondary are then modeled as spherical self-gravitating granular aggregates. However, cohesion is ignored in our investigation. The entire evolution process during the encounter is explored by using the soft-sphere implementation in \emph{pkdgrav}. In the model, to balance between computational efficiency and the ability to capture enough configuration details, the primary and secondary are given to be composed of 5616 and 1601 equal-size spherical particles, respectively, which are arranged randomly.

Previous simulations using pkdgrav conducted under a wide range of environments have demonstrated that the macroscopic friction of the simulated granular material can be controlled by the parameter set, ($\beta$, $\mu_s$, $\mu_r$, $\mu_t$) \citep{zhang2017creep, zhang2018rotational, schwartz2018catastrophic}. As suggested in \cite{jiang2015novel} and \cite{zhang2017creep}, these parameters are adopted in this study: $\beta$ = 0.5, $\mu_s$ = 0.5, $\mu_r$ = 1.05, and $\mu_t$ = 1.3, which give a moderate friction angle around 33$^\circ$ for a dense-random-packing rubble-pile body as evaluated by \cite{zhang2018rotational}\footnote{Although the macroscopic friction angle of a granular material is sensitive to the loading scenario and the particle arrangement, the actual friction angle of the rubble-pile model used in this study should be close to 33$^\circ$ given the similarity between our models and the models used in \cite{zhang2018rotational}.}. The normal and tangential coefficients of restitution, $\varepsilon_n$ and $\varepsilon_t$ are set to be 0.55 so that the granular system is subject to sufficient damping \citep{Chau2002coefficient}. The normal stiffness $k_n$ is determined by limiting the maximum particle overlaps not exceed 1\% of the smallest particle radius and the tangential stiffness $k_s$ is often set to (2/7)$k_n$ (see \cite{schwartz2012implementation} for details). We have verified that this set of values is able to sustain the current configuration of Toutatis modeled as cohesionless granular aggregates under the interactions between self-gravity and contact forces rotating in the current non-principal states. Herein we will not present the dependency of the newly formed contact binaries' configurations on these parameters. However, several simulations with variational parameters have shown that their variations do not make a major difference in the outcomes and  do not change our principal conclusions.

\subsection{Dynamical model}
The mutual orbit of a binary asteroid may be greatly perturbed by Earth's gravity during its encounter, and its fate after that depends on how much the semi-major axis and eccentricity are changed \citep{fang2011binary}. The secondary can escape from the binary system, or impact on the primary, even constitute an intact, stable binary system with the primary but at a different eccentricity. If the components impact each other, the secondary may be partially or fully accreted to create a contact binary. However, in our scenario we are mainly concerned with the latter outcome, in which no material escapes after impact. Also, bear it in mind that the tidal force of Earth may change the shape and rotation of the primary and secondary, if the approaching distance is small enough, which, combined with impact, will finally determine the outcomes.

Since the material properties have been set, the final results after encounter are controlled by 14 parameters: 6 Kepler elements of the mutual orbit, 6 parameters related to the spin state of the primary and secondary, the impact parameter $r_b$ (a hypothetical encounter distance that would result in the absence of Earth's gravity) and the hyperbolic encounter speed $v_\infty$, in which $r_b$ and $v_\infty$ are related to encounter distance $r_p$ and encounter speed $v_p$ at periapsis by
\begin{equation}\label{eq:rp_vp}
{r_b} = {r_p}\sqrt {1 + \frac{{2G{m_e}}}{{{r_p}v_\infty ^2}}} ,\;\;{v_\infty } = \sqrt {v_p^2 - \frac{{2G{m_e}}}{{{r_p}}}}
\end{equation}
where $G$ is the gravitational constant and $m_e$ is the mass of the Earth.

However, in view of the binary components' high mass ratio $q$ (up to 0.28), current observations of binary asteroids imply that the assumed precursor of Toutatis may be reasonably treated as a doubly synchronous binary with a tight mutual orbit \citep{pravec2007binary, walsh2015formation}. Accordingly, parameters can be reduced greatly by taking into account the doubly synchronous constraint. These assumptions are accordingly adopted in this study: orbital eccentricity $e = 0$, and the components' rotation periods both synchronous with the orbit period and their rotational orientations aligned with the normal direction of the mutual orbit. In addition, we will set the semi-major axis at a typical value $a = 4 R_p$ (corresponding to an orbit period of 16.17 hours) and $v_\infty = 9$ km/s (this is the peak value in the probability distribution curve of $v_\infty$ given by \cite{fang2011binary}) to reduce the complexity, but not change our conclusions in this work. Now the exploration parameter spaces are reduced to be dependent on 4 parameters: the inclination $i$, longitude of ascending node $\Omega$, $u = \omega + \nu$ (where $\omega$ is argument of periapsis and $\nu$ true anomaly) and $r_b$ (or $r_p$).

\begin{figure*}
\centering
\includegraphics[width=0.70\textwidth]{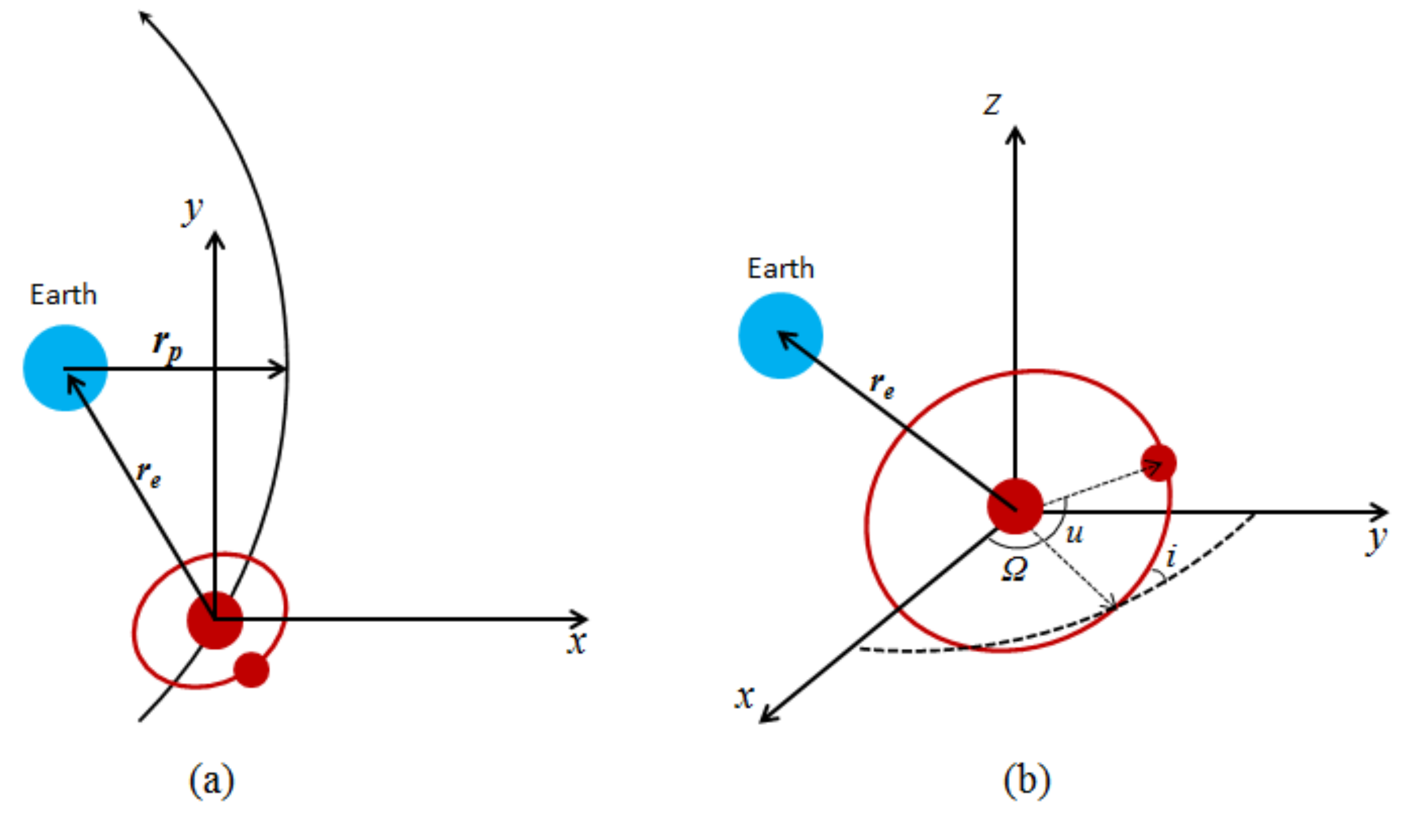}
\caption{Panel (a) and (b) are illustrations of the reference system used in our simulations. The coordinate origin is located at the center of the primary, the \emph{x}-axis is parallel to the vector from Earth to the periapsis of the hyperbolic orbit ($\bf{r_p}$), and the \emph{z}-axis is chosen as the normal direction of the flyby orbit. $\bf{r_e}$ is the position vector of Earth relative to the primary. Panels (a) and (b) are viewed in the \emph{z} direction and an arbitrary direction, respectively.}
\label{fig:reference}
\end{figure*}

If we take the components as rigid bodies and set the origin of reference system as the center of the primary (the reference system we adopt is shown in panels (a) and (b) of Fig. \ref{fig:reference}), then the motion of the secondary and Earth before impact is independent of their spin state. The dynamical equation of the orbit is

\begin{equation}\label{eq:orbit_equation}
\left\{ \begin{array}{l}
{\bf{\ddot r}} =  - \frac{{\left( {{\mu _1} + {\mu _2}} \right)}}{{{r^3}}}{\bf{r}} + {\mu _e}\left( {\frac{{{\bf{r}}_e} - {\bf{r}}}{{|{{\bf{r}}_e} - {\bf{r}}|^3}} - \frac{{{{\bf{r}}_e}}}{{r_e^3}}} \right)\\
{{{\bf{\ddot r}}}_e} =  - \frac{{\left( {{\mu _1} + {\mu _e}} \right)}}{{r_e^3}}{{\bf{r}}_e} - {\mu _2}\left( {\frac{{{\bf{r}}_e} - {\bf{r}}}{{|{{\bf{r}}_e} - {\bf{r}}|^3}} + \frac{{\bf{r}}}{{{r^3}}}} \right)\\
{\mu _1} = G{m_1},{\mu _2} = G{m_2},{\mu _e} = G{m_e}\\
r = |{\bf{r}}|,\;{r_e} = |{{\bf{r}}_e}|
\end{array} \right.
\end{equation}
where $m_1$, $m_2$, $\bf{r}$, and $\bf{r}_e$ are the masses of the primary and secondary, and the position vectors of the secondary and Earth, respectively. Solar gravitational perturbation is ignored in our model.

\begin{figure*}
\centering
\includegraphics[width=0.50\textwidth]{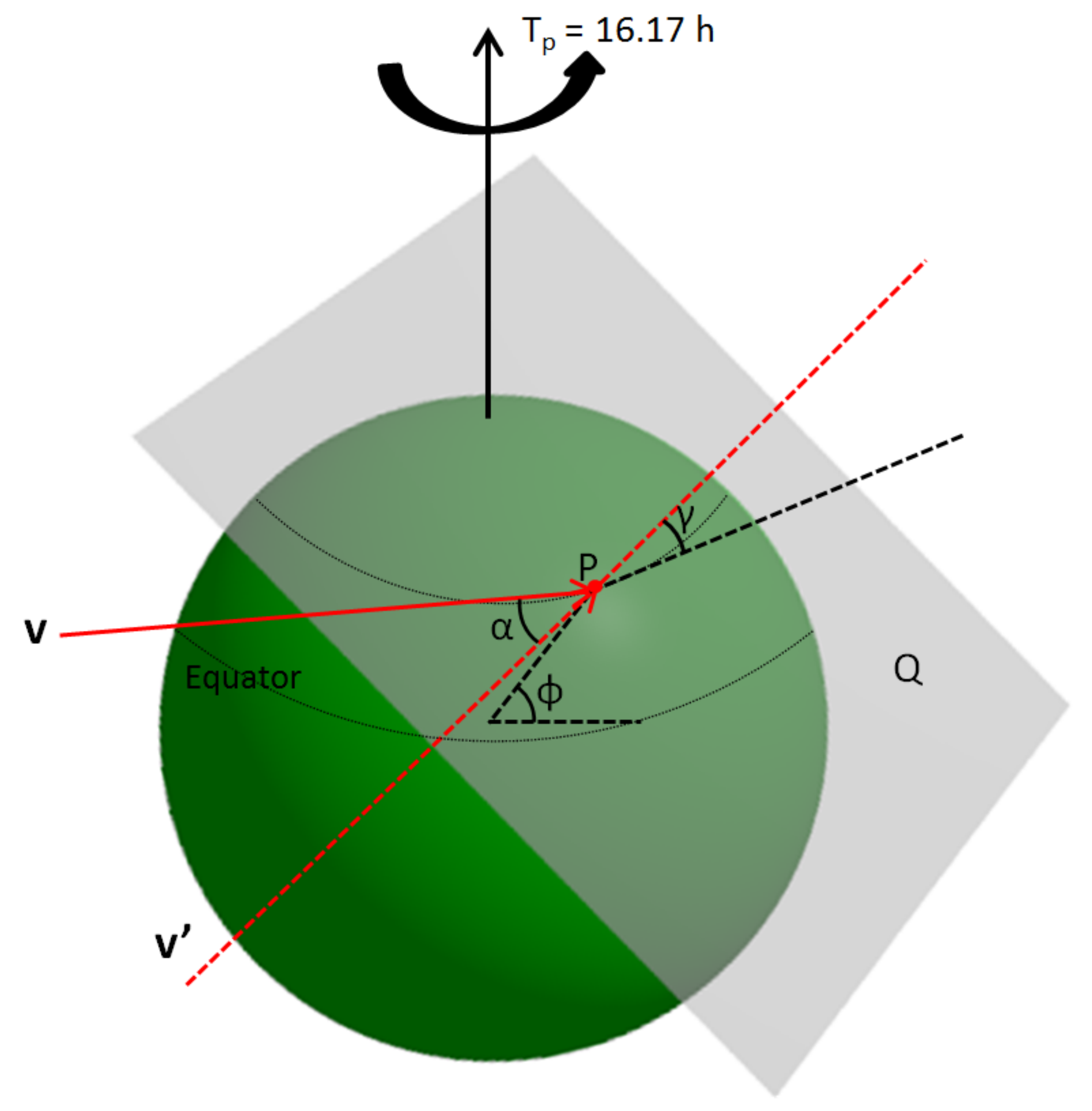}
\caption{An illustration of the four parameters related to impact in a rigid approximation: impact speed $v = |\bf{v}|$, impact angle $\alpha$, latitude of impact point $\phi$ and azimuth of impact velocity $\gamma$. The green sphere represents the primary (the secondary is not drawn). Impact velocity $\bf{v}$ is defined to be the relative velocity of the secondary relative to the primary. The rotation period of the primary $T_p$ is 16.17 hours.}
\label{fig:impact_parameters}
\end{figure*}

For the case of impact, the mutual orbit may be torqued by Earth's gravitational perturbation during encounter. As a result, the impact velocity may not lie in the equatorial plane of the primary. If we assume the rotations and shapes of two components remain unchanged in the process (rigid approximation), an illustration of the geometrical relationship at the impact moment is shown in Fig. \ref{fig:impact_parameters} (the secondary is not drawn). P is the impact point with latitude $\phi$, Q is the tangent plane on the surface of the primary that passes through P. $\bf{v}$ is the impact velocity and $\bf{v}'$ is the projection of $\bf{v}$ on plane Q. $\alpha$ is the impact angle, which is defined to be the angle between $\bf{v}$ and $\bf{v}'$. $\gamma$ is the azimuth of the impact velocity, which is the angle between $\bf{v}'$ and the local latitude line. These four parameters, $v = |\bf{v}|, \alpha, \phi$, and $\gamma$ are more intuitively related to the impact than the four initial parameters.

\begin{figure*}
\centering
\includegraphics[width=1.0\textwidth]{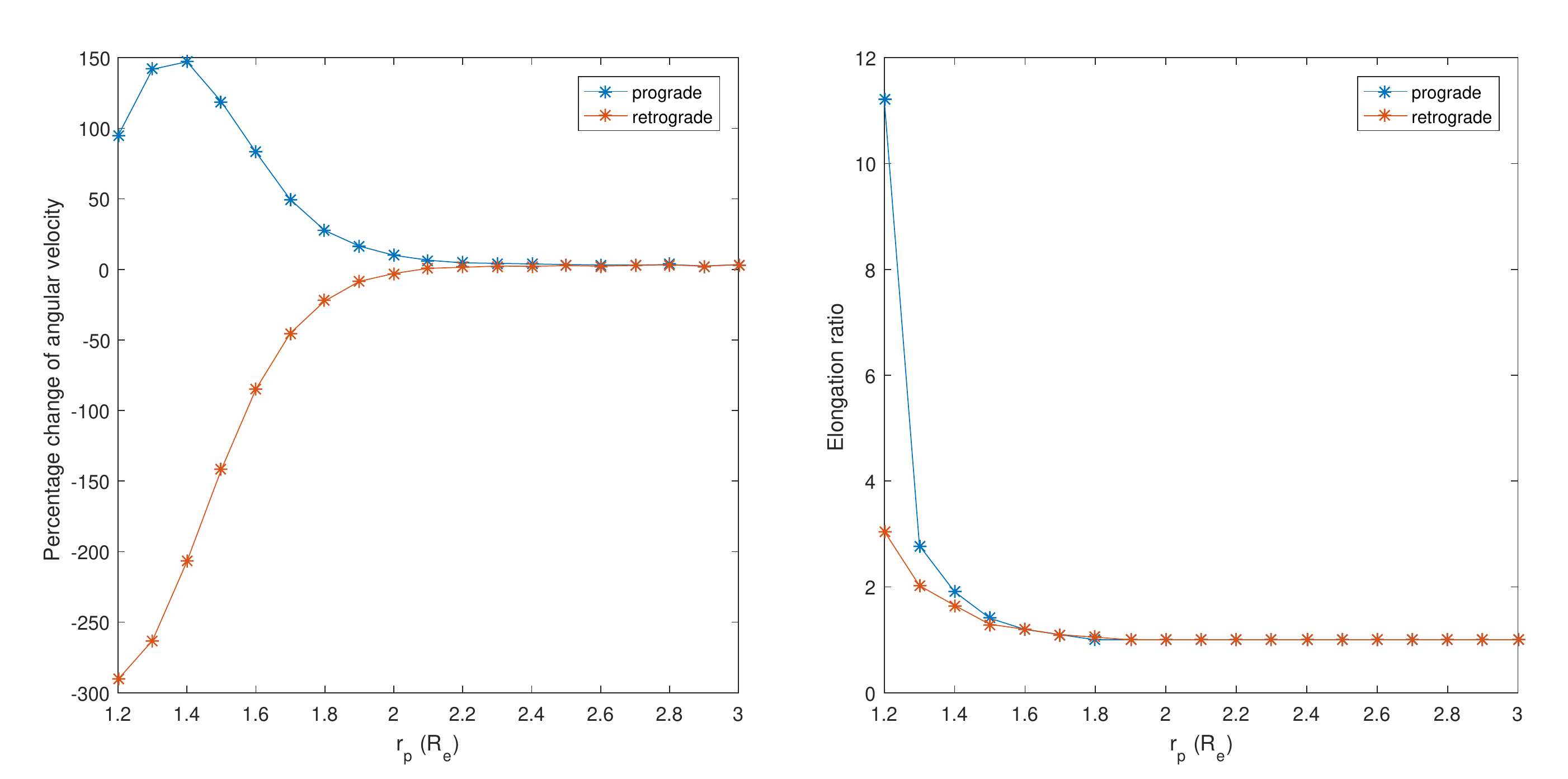}
\caption{Variation of (a) rotational angular velocity and (b) elongation ratio of the stable configuration after a close Earth encounter vs.\ the encounter distance. The blue lines correspond to an initial orbit with inclination $0^\circ$ (prograde) while red lines correspond to inclination $180^\circ$ (retrograde).}
\label{fig:body_head_change}
\end{figure*}

However, the rigid approximation is not accurate enough to reflect the real process. In this study, two different cases will be considered separately according to how much the Earth's tide will alter the components' shape and rotation in the process: (a) the tidal effect can be ignored; (b) the tidal effect cannot be ignored, which depends on the encounter distance $r_p$. For (a), the merging process may be regarded as only subject to the physics of the impact. But for (b), the situation is much more complicated as shape reconfiguration, rotation change, and impact are coupled with the Earth's tide, which makes results be more diverse and sensitive to initial conditions. Therefore, we carried out some numerical experiments simply using the primary (without the secondary) as the progenitor to roughly estimate the critical encounter distance $r_c$ with \emph{pkdgrav}. Fixing the rotational period of the spherical progenitor to be the same as the mutual orbit period at $a = 4 R_p$, we considered two extreme rotational orientations: prograde and retrograde spin relative to the flyby orbit. The progenitor experiences a close Earth encounter process just as the binary does, in which $r_p$ is varied from 1.2 $R_e$ to 3.0 $R_e$ with an increment of 0.1 $R_e$. The change of rotational angular velocity ($\omega$) and resulting elongation ratio ($\epsilon$) of the progenitor after encounter are calculated and presented in Fig. \ref{fig:body_head_change}, which shows that the variation of $\omega$ may be well constrained below 5 \% and $\epsilon$ is almost not changed at $r_p > 2.5 R_e$. However, the tiny changes of rotation and shape at $r_p > 2.5 R_e$ have little influence on the outcomes. Accordingly, as will be shown hereafter, our methodology will be different in these two situations: $r_p > 2.5 R_e$ and $r_p < 2.5 R_e$.

\subsection{Monte Carlo simulations}
Since the \emph{N}-body simulations are computationally expensive and the parameter space to explore is very large, we initially use a Monte Carlo method \citep{fang2011binary, fang2011orbits}. For our exploration, $u$ is drawn from a uniform distribution, the normal directions of mutual orbits are selected in an isotropic way (then $i$ and $\Omega$ are determined), and the occurrence probability of $r_b$ is $k\cdot r_b$ ($k$ is a constant).

\begin{figure*}
\centering
\includegraphics[width=0.8\textwidth]{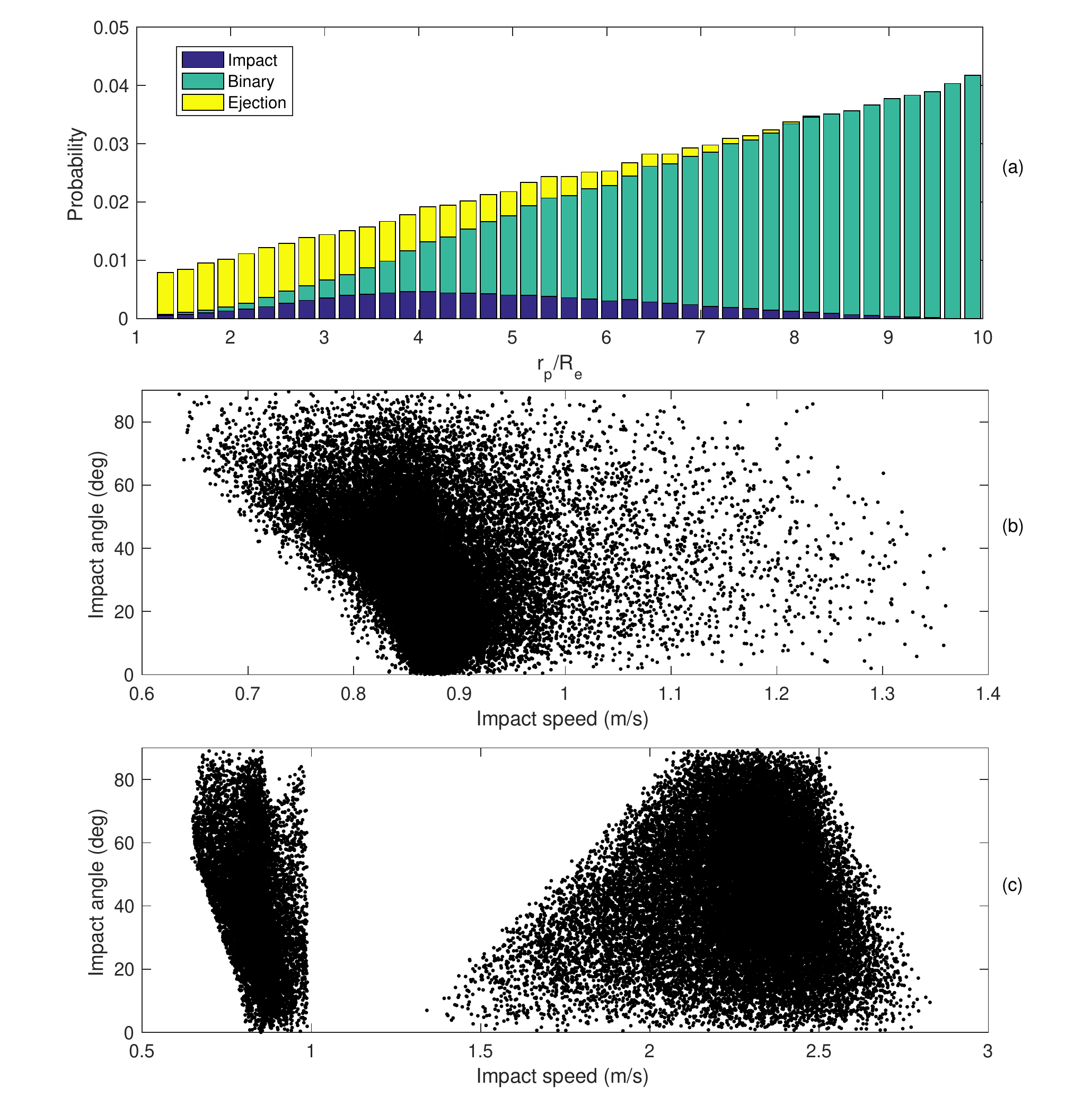}

\caption{Monte Carlo simulation results of the mutual orbits. Panel (a): probabilities of three fates of the binary system vs. $r_p$ (500,000 simulations were performed and 9.6\% of cases in total experienced mutual impact). Panel (b): scatter plot of $\alpha$ vs. $v$ for $r_p > 2.5 R_e$. Panel (c): scatter plot of $\alpha$ vs. $v$ for $1.4 R_e \leq r_p \leq 1.5 R_e$. 40,000 impact cases were run and are shown in panels (b) and (c). The great gap between 1 m/s $ < v < $ 1.3 m/s in panel (c) is because we simply consider the inbound-type hyperbolic orbits. }
\label{fig:distribution_vs_angles}
\end{figure*}

We  numerically performed orbital integrations by randomly walking through the parameter space with the given probability distributions. We noted 3 outcomes (fates) in these simulations: impact, binary, and ejection. Probabilities of three fates vs. encounter distance $r_p$ are displayed in the panel (a) of Fig. \ref{fig:distribution_vs_angles} ($1.2 R_e < r_p < 10 R_e$). The results show that mutual impacts are possible for $r_p < 10 R_e$ and the probability reaches a maximum value for $r_p \sim 4 R_e$. Impact speed $v$ and impact angle $\alpha$ are obtained from the impact cases and the scatter plot for $r_p > 2.5 R_e$ and $1.4 R_e \leq r_p \leq 1.5 R_e$ are shown respectively, in panels (b) and (c) of Fig. \ref{fig:distribution_vs_angles}.

By analyzing the simulation results in the panel (a) of Fig. \ref{fig:distribution_vs_angles}, we notice that an interesting characteristic that all impact cases experienced collisions after passing the perigee in our model, indicating the impacts take place in the outbound flyby orbits no matter how the rotation and shape of each component is altered by Earth's tides. Denoting $r_i$ as the distance from Earth to the binary at the moment of impact, we find that 100\% of impact cases occur for $r_i > 2 R_e$, whereas 99.8\% for $r_i > 2.5 R_e$, 96.7\% for $r_i > 5 R_e$ and 85.8\% for $r_i > 10 R_e$. Combined with the critical $r_c = 2.5 R_e$, we can infer that, in most cases, the rotation and shape of each component have finished their modifications due to Earth's tide prior to impact, and the tidal distortion will not be coupled with the subsequent impact process.

\section{Results}
\subsection{$r_p > 2.5 R_e$}
In this case, the gravity of Earth will change the mutual orbit of the system during the encounter, while the influence on the shape and rotation can be safely ignored. Then the subsequent evolution after impact is only subject to the four parameters $v, \alpha, \phi$, and $\gamma$.

\begin{figure*}
\centering
\includegraphics[width=1.0\textwidth]{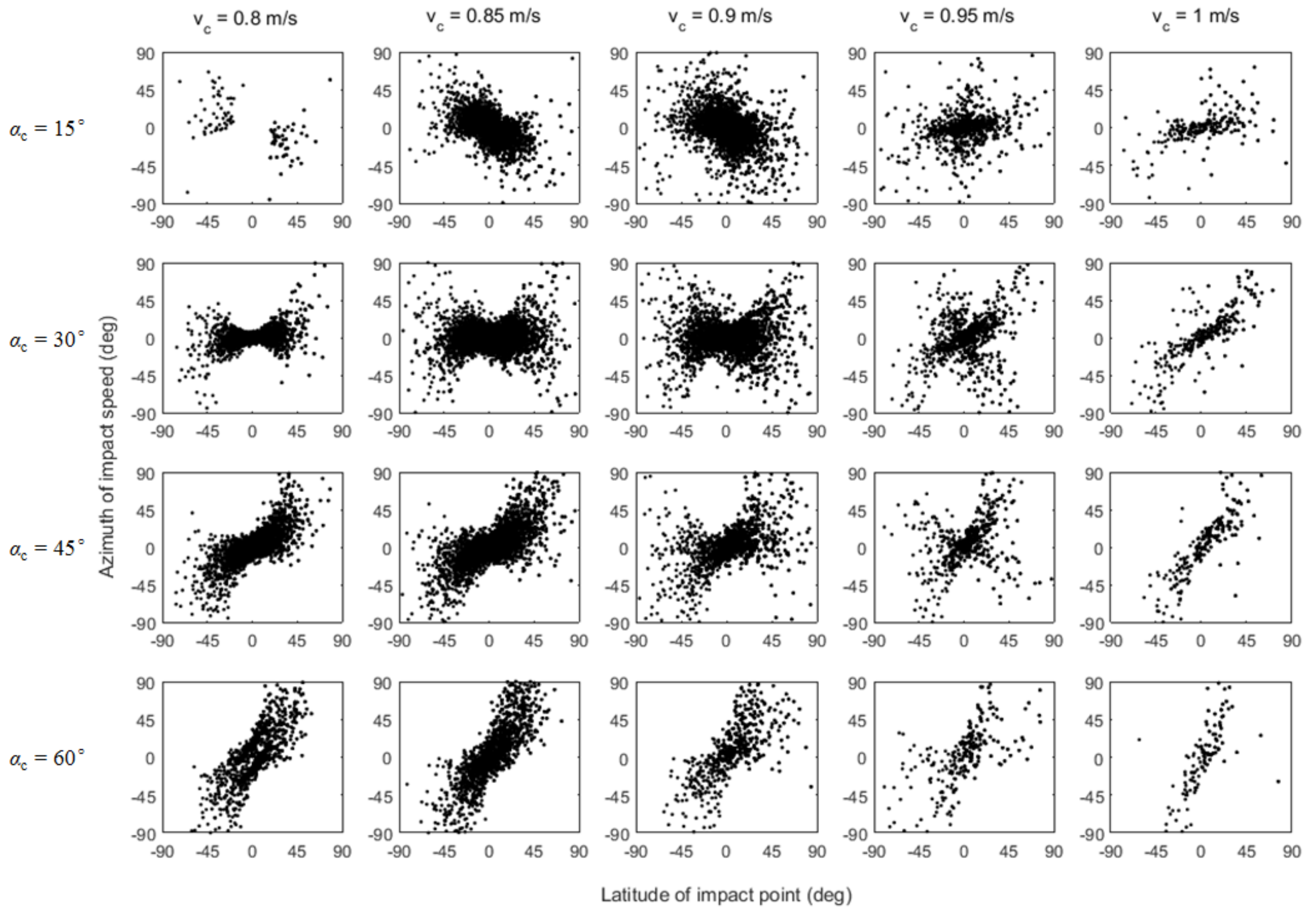}
\caption{Scatter plot of azimuth of impact speed $\gamma$ vs. latitude of impact point $\phi$ in intervals $|v-v_c| < 0.025$ m/s and $|\alpha-\alpha_c| < 7.5^\circ$. $v_c$ and $\alpha_c$ are selected as $0.8, 0.85, 0.9, 0.95, 1.0$ m/s, and $15^\circ, 30^\circ, 45^\circ, 60^\circ$, respectively. The scattering pattern in each panel is centrosymmetric at $\phi = 0^\circ$ and $\gamma = 0^\circ$. The data is captured from the results in the panel (b) of Fig. \ref{fig:distribution_vs_angles}.}
\label{fig:distribution_vs_anglesx}
\end{figure*}

A scatter plot of  $\alpha$ vs. $v$ for $r_p > 2.5 R_e$ is displayed in the panel (b) of Fig. \ref{fig:distribution_vs_angles}, from which we can see that the points are not distributed uniformly. About 96.1\% cases have $v < 1.0$ m/s (1 m/s is the escape speed at $r = R_p + R_s$ in two-body problem), which implies mutually merging or orbiting are likely to happen after impact. In order to get more insight into the distribution of the four impact parameters, scatter plots of $\phi$ vs. $\gamma$ are shown separately in 20 intervals ($|v-v_c| < 0.025$ m/s and $|\alpha-\alpha_c| < 7.5^\circ$) in Fig. \ref{fig:distribution_vs_anglesx}. The results show that low $|\phi|$ and low $\gamma$ dominate in the distributions and $\gamma$ tends to distribute more widely as $|\phi|$ increases.

\begin{figure*}
\includegraphics[width=1.00\textwidth]{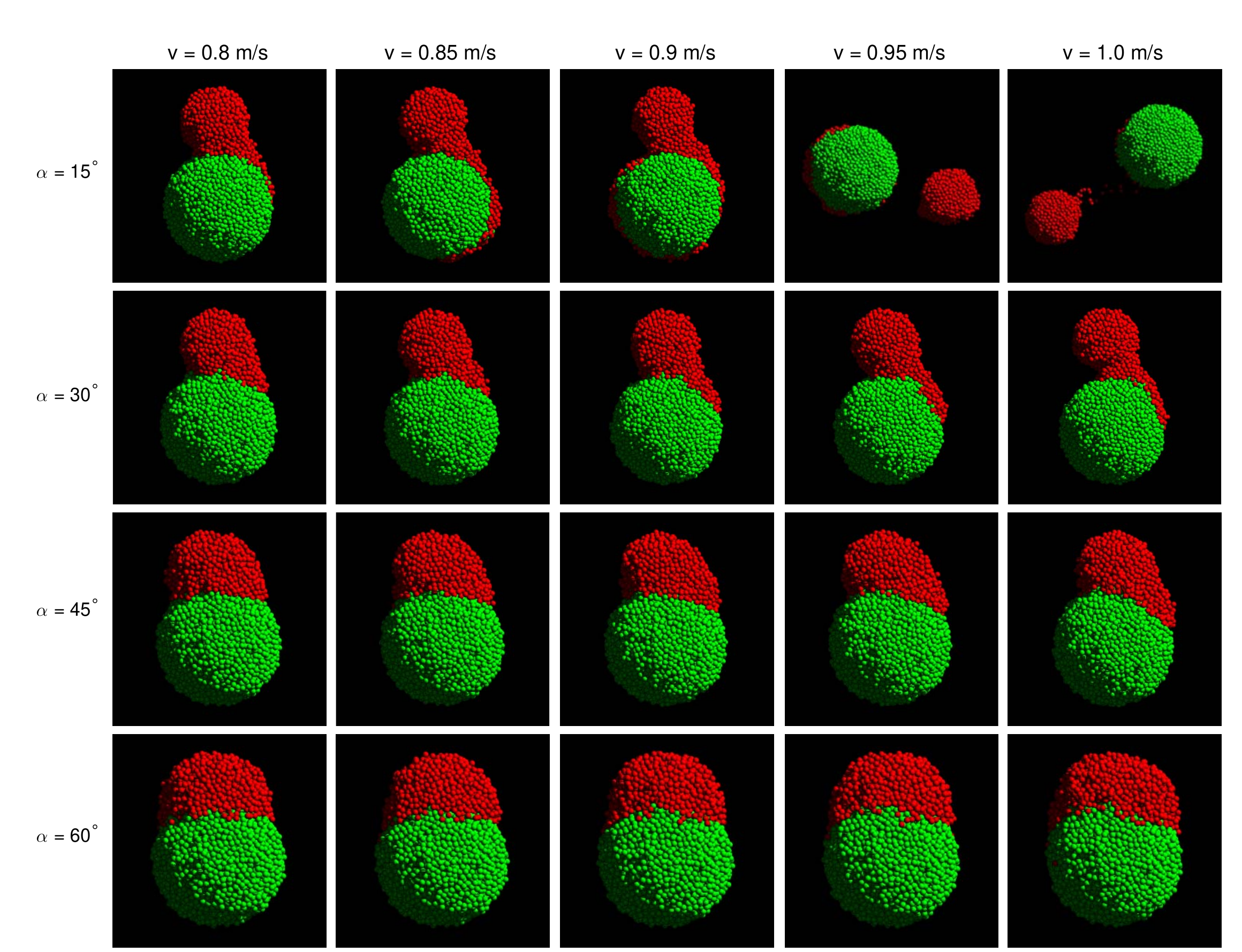}
\caption{The configurations of newly formed contact binaries for $\alpha = 0^\circ$ and $\gamma = 0^\circ$. $v$ is varied from 0.8 m/s to 1.0 m/s with increment 0.05 m/s and $\alpha$ is varied from $15^\circ$ to $60^\circ$ with increment $15^\circ$.}
\label{fig:impact_result_equator}
\end{figure*}

 For $\phi = 0^\circ, \gamma = 0^\circ$, we set $v = 0.8, 0.85, 0.9, 0.95, 1.0$ m/s and $\alpha = 15^\circ, 30^\circ, 45^\circ, 60^\circ$, and used \emph{pkdgrav} to simulate the subsequent evolution. The outcomes of 20 cases are shown in Fig. \ref{fig:impact_result_equator}, in which integrations were performed for 10 simulated days to obtain stable configurations. We can see that, 18 out of 20 cases are fully accreted, in each of which the binary system is merged and turns into a contact binary. However, for [$v$ = 0.9 m/s, $\alpha = 15^\circ$] and [$v$ = 1.0 m/s, $\alpha = 15^\circ$], some part of the secondary's material is accreted whereas most remains in orbit around the primary. After impact, the secondary begins to move with velocity components $v\sin\alpha$ and $v\cos\alpha$ in the radial and horizontal directions, respectively (ignoring the spin of each component). Both the radial and horizontal movement will halt until friction between materials dissipates enough kinetic energy. Generally, it is found that collapse is enhanced as $\alpha$ increases, and more materials of the secondary are scattered along the movement path on the surface of the primary as $v$ increases, because it takes more time for the secondary to reduce its speed in that case. For $\alpha = 60^\circ$, the boundary between the primary (body) and secondary (head) is so blurred that the characteristic of a contact binary is not apparent.

\begin{figure*}
\includegraphics[width=1.00\textwidth]{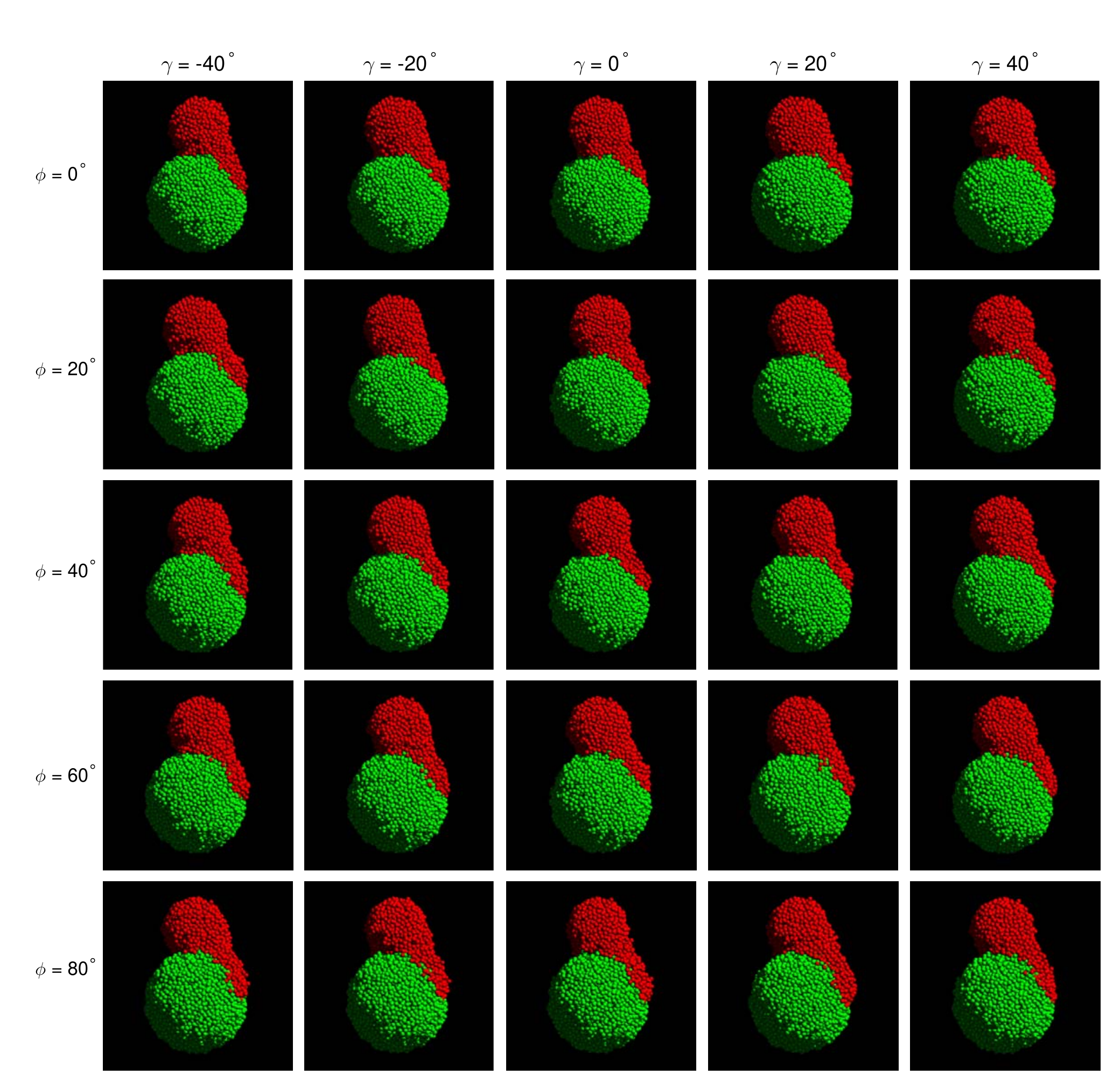}
\caption{{The configurations of newly formed contact binaries for $v$ = 0.9 m/s and $\phi = 30^\circ$. $\phi$ is varied from $0^\circ$ to $80^\circ$ with increment $20^\circ$ and $\gamma$ is varied from $-40^\circ$ to $40^\circ$  with increment $20^\circ$.}}
\label{fig:impact_result_LA}
\end{figure*}

For $\alpha \neq 0^\circ$ and $\gamma \neq 0^\circ$, however, the direction of the orbit angular momentum ($\bf{L}_{orb}$) does not align with the rotational angular momentum ($\bf{L}_{rot}$). The horizontal movement of the secondary materials on the primary is affected by $\phi$ and $\gamma$. Taking the case [$v = 0.9$ m/s, $\alpha = 30^\circ$] in Fig. \ref{fig:impact_result_equator} as a nominal example, we have expanded $\phi = 0^\circ$ to $\phi = 0^\circ, 20^\circ, 40^\circ, 60^\circ, 80^\circ$ and $\gamma = 0^\circ$ to $\gamma = -40^\circ, -20^\circ, 0^\circ, 20^\circ, 40^\circ$ respectively, based on the distribution in Fig. \ref{fig:distribution_vs_angles}. Fig. \ref{fig:impact_result_LA} shows 25 simulation results among our runs. Note that the variations of $\phi$ and $\gamma$ do not make major differences in the outcomes. Actually, for [$v$ = 0.9 m/s, $\alpha = 30^\circ$], the rotational velocities of the primary and secondary at the equator are 0.12 m/s and 0.08 m/s, respectively, both of which are much less than the centroidal horizontal velocity 0.78 m/s. All results suggest that, in our model, the outcomes of mutual impacts in this situation are dependent primarily on $v$ and $\alpha$, but not so significantly rely on $\phi$ and $\gamma$.

Moreover, we further notice that distinct contact binary configurations are evident in Fig. \ref{fig:impact_result_equator} (such as [$v = 0.8$ m/s, $\alpha = 15^\circ$], [$v = 0.85$ m/s, $\alpha = 30^\circ$], [$v = 0.9$ m/s, $\alpha = 30^\circ$] and [$v = 0.95$ m/s, $\alpha = 30^\circ$]) and Fig. \ref{fig:impact_result_LA}. However, the body shapes remain still approximately spherical. These results demonstrate that the low-speed impact of the secondary has little effect on the shape of the primary while the shape of the secondary is strongly deformed.

\subsection{$r_p < 2.5 R_e$}
For $r_p < 2.5 R_e$, tidal effects arising from Earth will change the rotation and shape of each component during the close approaches. However, the results for $r_p > 2.5 R_e$ suggest that the low-speed impact may not have an influence on the body elongation. Based on the outcomes in the right panel of Fig. \ref{fig:body_head_change}, we expect to constrain the flyby distance $r_p$ to be in the range $1.4 - 1.5 R_e$ , to acquire a satisfactory elongation ratio of the primary to match the optical images seized from the Chang'e-2's flyby.

In this case, the mutual orbit after flyby may be strongly perturbed by the combined non-spherical gravity of the primary and secondary resulting from their reshaped configurations. Nevertheless, for an approximation, we have additionally calculated $v$ and $\alpha$ in the same way as we did in Section 3.1. The scatter plot is shown in the panel (c) of Fig. \ref{fig:distribution_vs_angles} for $1.4 \leq r_p \leq 1.5 R_e$. We observe that the points in the figure are clearly separated into two parts: $v < 1$ m/s (left branch, occupying 26.2\%) and $v > 1$ m/s (right branch, for 73.8\%). Apparently, the left branch and right branch correspond to elliptical and hyperbolic impact orbits, respectively, indicating that the initial conditions in the right branch is more likely to result in "touch and go" than those of the left branch.

As a result, we simply consider the initial conditions from the left branch. In addition, the outcomes for $r_p > 2.5 R_e$ imply that a lower impact angle is more likely to result in a distinct contact binary. In view of this, we further constrain $\alpha < 45^\circ$ and then 128 groups of the remaining initial parameters are selected randomly. The outcomes show that only 87 cases merge together and create contact binaries without material loss. Compared with the results shown in Fig. \ref{fig:impact_result_equator} and \ref{fig:impact_result_LA}, the configurations in this case are more diverse, and can be accordingly classified into three types on the basis of the location of the contact point between two components:

(A) Contact point along the long axis of the primary (12 cases  out of 87).

(B) Contact point along the short axis of the primary (39 cases out of 87).

(C) Resulting configurations between (A) and (B) (36 cases out of 87).

\begin{figure*}
\includegraphics[width=0.80\textwidth]{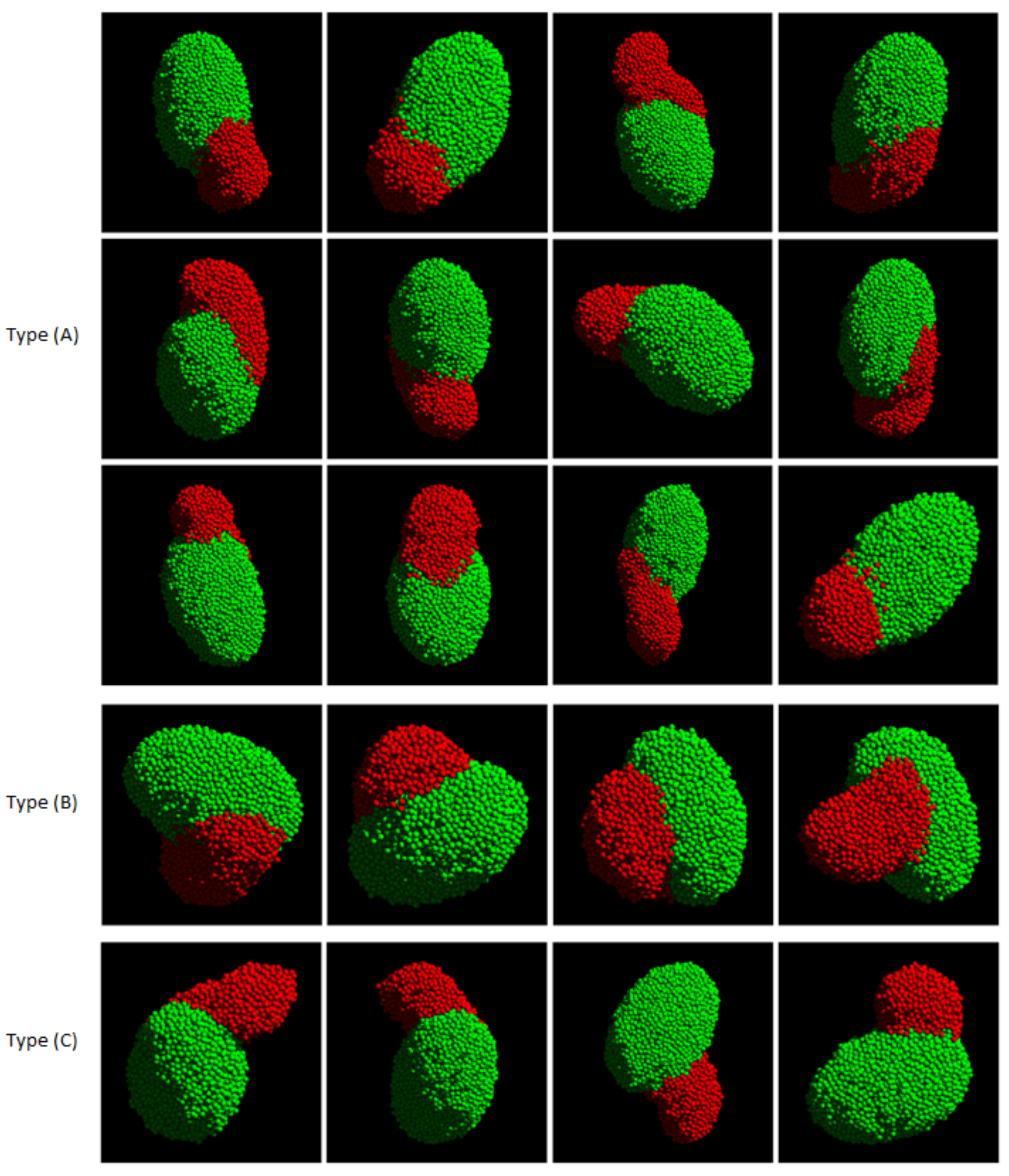}
\caption{Illustrations of the three types of final stable configurations of contact binaries formed after impact for $1.4 R_e \leq r_p \leq 1.5 R_e$. Initial conditions are randomly selected from the left branch of the panel (c) in Fig. \ref{fig:distribution_vs_angles} with a constraint of $\alpha < 45^\circ$.}
\label{fig:tidal_many_pic}
\end{figure*}

Fig. \ref{fig:tidal_many_pic} summarizes our results, where 12 newly formed type (A) contact binary cases are shown in the top panel, and 4 typical cases of type (B) and (C) are presented in the middle and bottom panel, respectively. Similar to those of Section 3.1, the mutual impacts do not have a significant influence on the overall shape of the final body.

The outcomes of type (B) appear to be more common than those of type (C). This is because such kind of configuration corresponds to the minimum gravitational potential, thereby remaining most stable amongst the simulations. Obviously, Toutatis-like cases can only be identified from type (A) runs, where we observe that the resulting configurations are diverse for various impact parameters. Most cases in type (A) have collapsed-head configurations and the separations of two components are not easy to be discerned. Several possible reasons are summarized as follows:

(1) The primary and secondary may spin up due to Earth's tidal torque, therefore the relative impact speed at the contact point can  be enhanced;

(2) The end of the long axis of the primary has a high gravitational potential and raised surface, which makes materials more likely to slide without the constraint of cohesion.

\begin{figure*}
\includegraphics[width=0.80\textwidth]{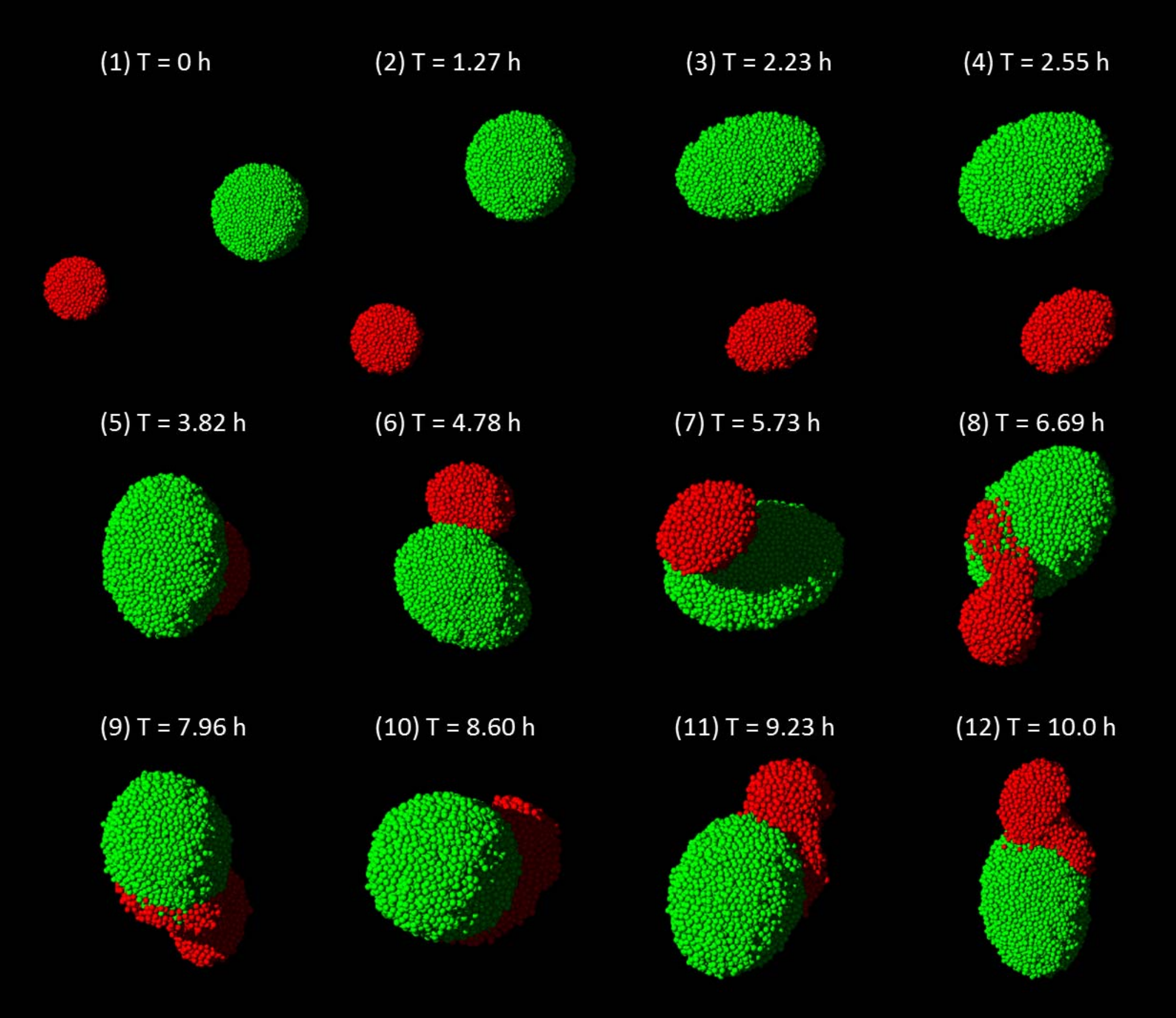}
\caption{Snapshots of the merging process of the binary system after a close Earth encounter ($r_p \sim 1.5 R_e$). In this case, both the shapes of primary and secondary are stretched by tidal force. The shape of the primary is almost not affected by the impact and a portion of the material of the secondary is scattered on the surface of the primary.}
\label{fig:tidal_bestfit}
\end{figure*}

Even for cases that happen to directly impact at the end of the primary's long axis, the movement of the secondary relative to the primary will not stop right after the impact, but likely continue until the secondary rolls down to lower terrain. However, we can also see that the third panel in the first row has a good Toutatis-like shape, in which a distinct contact binary configuration and elongated body are obtained. The evolution process of this run is in details shown in Fig. \ref{fig:tidal_bestfit}. We can clearly notice that the shapes of the primary and secondary are stretched out due to Earth's tidal effect before impact. The secondary first impacts on a lower-terrain area of the primary and rolls on its surface inertially and happens to stop at the end of the long axis once friction dissipates enough energy. Some materials of the secondary are lost and scattered along the movement path. The shape of the elongated body is preserved during the low-speed impact.

However, our simulations suggest that it is not easy to obtain a satisfactory result for this situation. For a conservative estimation, after imposing a rigorous constraint on $r_p$ ($1.4 R_e < r_p < 1.5 R_e$), and selecting samples simply from the left branch of impact cases in the panel (c) of Fig. \ref{fig:distribution_vs_angles} with a constraint of $\alpha < 45^\circ$, the likelihood of forming an elongated contact binary is 1/87 - 12/87 (which means that we take into account all type (A) outcomes). Since we only considered a narrow parameter space (where we fixed $v_\infty$ and $a$, and assumed a doubly synchronous rotation state), the number of simulations that we explored using \emph{pkdgrav} is very limited, the estimated probability seems to be relatively rough. In our future study, we will extensively estimate the role of tidal distortion of involving terrestrial planets and investigate more diverse contact binary configurations.

Nevertheless, as aforementioned, Fig. \ref{fig:tidal_bestfit} shows the time evolution of two components of the binary system when they make close approaches to Earth. Herein the final results can reproduce a best-matching configuration as compared with those images of Toutatis by Chang'e-2 flyby mission. The scenario indicates that Earth's tides (or tidal disruption arising from other terrestrial planet) can reshape a binary system by make two components first being elongated, then experience colliding with each other at mutual low-speed, finally forming a bifurcated Toutatis-like elongated binary along with the end of long axis of the primary, thereby serving as a likely formation mechanism of Toutatis (or similar small bodies).

\subsection{Rotation analysis}
In this study, we do not adopt nowadays spin state of Toutatis as a constraint on the formation of contact binary configurations. However, in order to fully understand the outcomes after impact, thus we explore the rotation evolution of the asteroid when  encountering Earth. Herein we mainly concentrate on the variations of the primary's spin state, which is consistent with the system once the final configuration reaches equilibrium. In our investigation, we take the following parameters into account -- the rotation period, longitude and latitude of orientation, the angle between the rotation axis and the long axis of the newly formed system.

\begin{figure*}
\includegraphics[width=1.00\textwidth]{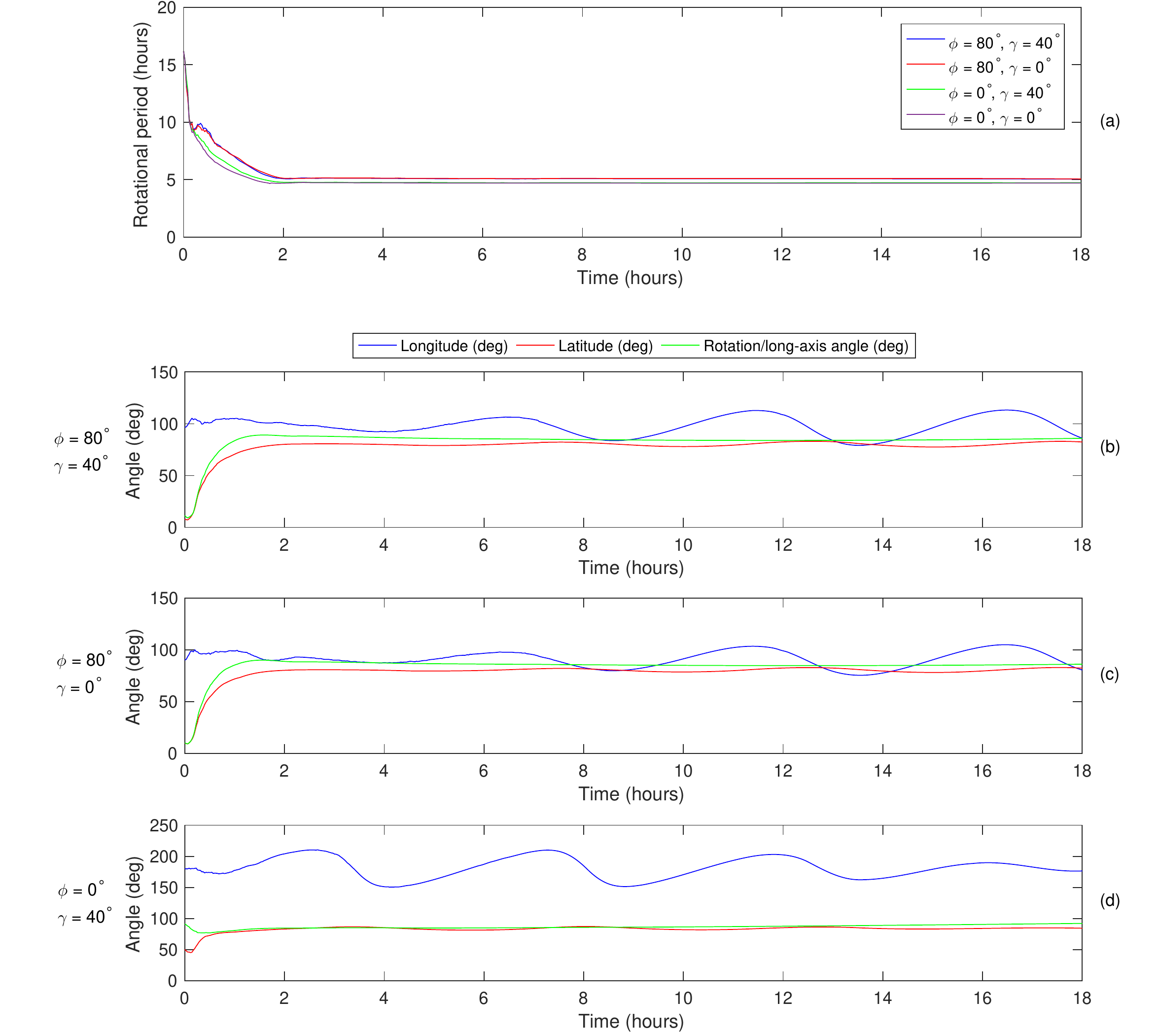}
\caption{Variation of the primary's rotational parameters with time after impact for $r_p > 2.5 R_e$. Panel (a) shows the variation of rotational periods for $[\phi = 80^\circ, \gamma = 40^\circ]$, $[\phi = 80^\circ, \gamma = 0^\circ]$, $[\phi = 0^\circ, \gamma = 40^\circ]$ and $[\phi = 0^\circ, \gamma = 0^\circ]$. The other three panels (b), (c) and (d) show the variation of orientation for $[\phi = 80^\circ, \gamma = 40^\circ]$, $[\phi = 80^\circ$, $\gamma = 0^\circ]$ and $[\phi = 0^\circ, \gamma = 40^\circ]$, in which the blue and red line in each panel are longitude and latitude of rotation orientation respectively, while the green line is the angle between the rotation axis and the long axis (where the long axis is approximated as the direction from the instantaneous centeroid of the primary (or body) to the secondary (or head)). For these three cases, the oscillations of orientation imply they are tumbling rotators. For $[\phi = 0^\circ, \gamma = 0^\circ]$, the variation of orientation is not given because the rotation is uniform.}
\label{fig:rotation_A}
\end{figure*}

For $r_p > 2.5 R_e$, Fig. \ref{fig:rotation_A} shows the variations of four parameters with time after impact for cases $[\phi = 80^\circ, \gamma = 40^\circ]$, $[\phi = 80^\circ, \gamma = 0^\circ]$, $[\phi = 0^\circ, \gamma = 40^\circ]$ and $[\phi = 0^\circ, \gamma = 0^\circ]$. The results indicate that rotation periods have been reduced from 16.17 hours to about 5 hours. For $v = 0.9 $ m/s, $\alpha = 30^\circ$, we have $|\bf{L}_{orb}| \approx$ 6$|\bf{L}_{rot}|$. Note that $\bf{L}_{orb}$ has been converted to $\bf{L}_{rot}$ during the merging scenario, which can explain the system's spin-up even though the moment inertia is increased. For $\phi = 0^\circ, \gamma = 0^\circ$, the rotation becomes uniform, with rotation axis perpendicular to the long principal axis. However, the rotations of the other three cases are tumbling, with precession periods of about 5 hours (estimated approximately from the oscillation of orientation curve). Moreover, the rotation/long-axis angles for three cases are $86.0^\circ$, $86.1^\circ$ and $92.0^\circ$, respectively, implying that the three contact binaries rotate almost perpendicular to their long axes but not along their long axes as Toutatis does currently. Compared to the actual spin state, these results suggest that the contact binaries formed in this way may have a totally different starting spin state. The main reason is that orbital angular momentum dominates in the merging process. The possible way to slow down the rotation is to increase the impact angle, so as to reduce $\bf{L}_{orb}$. However, in our cohesionless model, the compression due to impact will make the system collapse, as we can see from the results for $\alpha = 60^\circ$ in Fig. \ref{fig:distribution_vs_anglesx}.

\begin{figure*}
\includegraphics[width=1.00\textwidth]{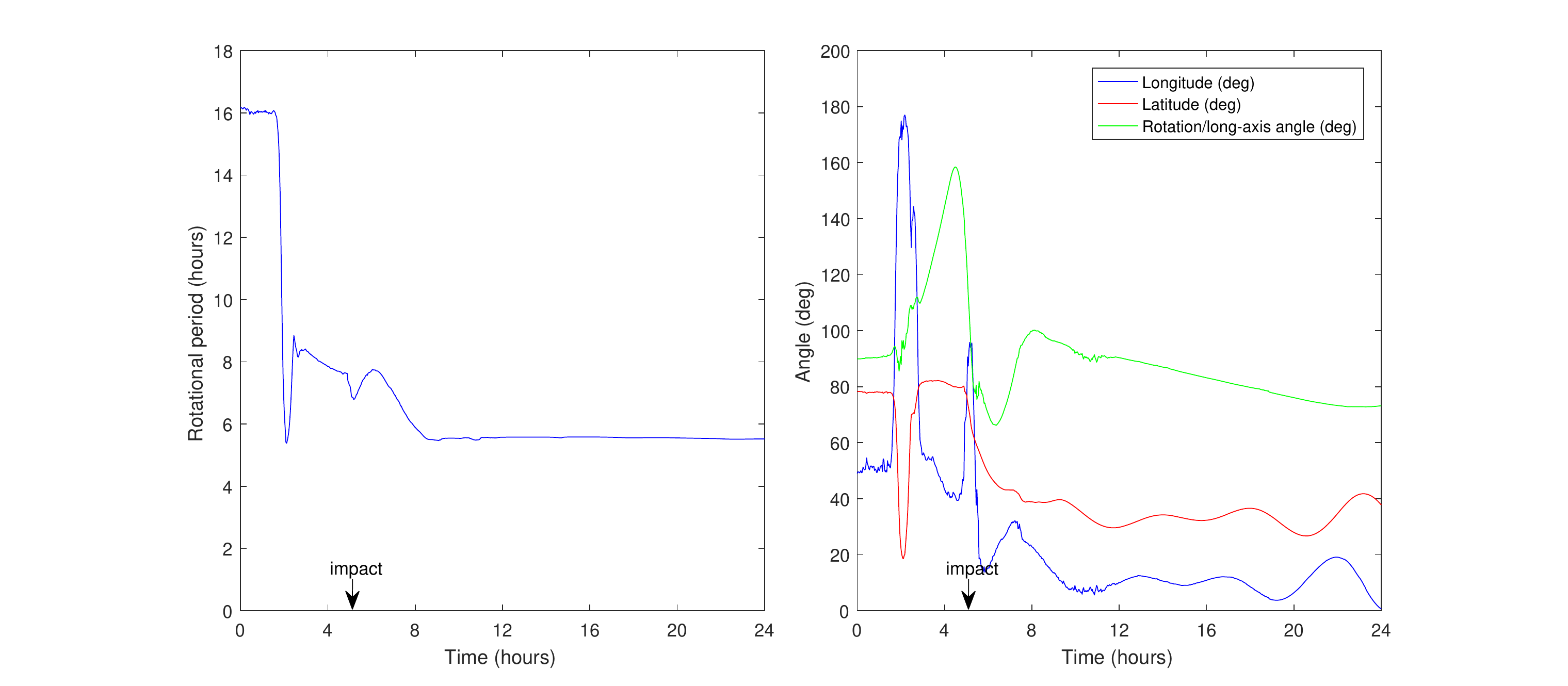}
\caption{Variation of the primary's rotational parameters before and after the impact for the case in Fig. \ref{fig:tidal_bestfit}. The left panel shows the rotational period, which decreases from 16.17 hours to 5.5 hours. The right panel shows the variation in orientation, for which the meaning of the three lines is the same as in Fig. \ref{fig:rotation_A}. An irregular tumbling rotation is also noticed in the right panel results, as we can see the oscillations of orientation (blue and red line).}
\label{fig:rotation_B}
\end{figure*}

For $r_p < 2.5 R_e$, we show similar profiles in Fig. \ref{fig:rotation_B} of the case in Fig. \ref{fig:tidal_bestfit}. The system can be spun up after Earth-approaching, with a rotation period decreasing from 16.17 hours down to 5.5 hours. The rotation axis is not along long axis, but has a large offset of 73$^\circ$. Note that tumbling rotation occurs in this case, with a precession period of about 5 hours. This result greatly differs from that of the actual spin state of Toutatis, similar to the case $r_p > 2.5 R_e$. Nevertheless, it appears to be possible that the rotation of the new system slows down in a complex way, in which $\bf{L}_{rot}$ is increased to a value comparable to $\bf{L}_{orb}$ by Earth's tides, while the mutual orbit switches from prograde to retrograde motion relative to the rotation, and then the opposite angular momentum direction makes most angular momentum be canceled out and form a slow tumbling rotator \citep{chauvineua1995evolution}. However, this situation has a very small chance of occurring and so far we have not found a case with a slow tumbling rotation similar to the actual state.

In summary, we find our distinct contact binaries tend to rotate along their short axes rather than the long axes since a low impact angle is preferred. Tumbling rotation is possible but the rotation and precession period in our model are much shorter than the actual values. As aforementioned, since an asteroid's rotation is easier to change than its shape, the mismatch between rotation states does not change our conclusion. Nevertheless, head-on impact (or impact with a high impact angle) may relieve the rotation problem. But cohesion would be needed in this case, since otherwise collapse would likely happen under this circumstance. The perpendicular angle about the neck region of Toutatis suggests the possible existence of cohesion. Detailed cohesion effects will be done in future investigation.

\section{Discussion and Conclusion}
4179 Toutatis is an interesting near-Earth asteroid with a bifurcated shape and unusual slow tumbling rotation. Optical images acquired by Chang'e-2 spacecraft suggest that the asteroid has a distinct bi-lobe structure with the contact point between the two lobes of body and head, located at the end of the body's long axis \citep{huang2013ginger}. The geological features of Toutatis \citep{hudson2003high,huang2013ginger} imply that this asteroid is not a monolith but most likely bears a rubble pile structure. Therefore, we speculate that the origin of Toutatis' configuration may be attributed to a low-speed impact event.

To examine this scenario and better understand the formation process of Toutatis, we carried out a series of numerical simulations by investigating the scenario that an assumed binary precursor performs a close Earth encounter leading to a low-speed impact between the primary and secondary. Monte Carlo simulations of the mutual orbit were performed to provide appropriate initial conditions for the computationally expensive \emph{N}-body simulations of the system, which were carried out by using the \emph{pkdgrav} package in a SSDEM implementation. Two components of the binary were represented as spherical cohesionless self-gravitating granular aggregates, whose physical parameters were obtained from our combined shape model of Toutatis obtained by synthesizing radar model and optical images. The mutual semi-major axis $a = 4 R_p$ and hyperbolic encounter speed $v_\infty = 9$ km/s were chosen to narrow down the parameter space.

We identified three major characteristics that our numerical models are required to match the Chang'e-2 flyby images: a distinct contact binary configuration, a body shape with elongation ratio $\sim 1.5$, and a contact point along the long axis of the body. The outcomes show that our scenario feasibly creates distinct contact binaries with given appropriate initial conditions. But the elongated body cannot be explained by mutual impact between the primary and secondary along. However, for $1.4 R_e \leq r_p \leq 1.5 R_e$, we showed a case that matches the aforementioned three characteristics very well, which suggests that a close terrestrial planet (Earth) encounter of a separated binary system can act as a likely formation mechanism, to produce contact configurations of Toutatis-like binaries, e.g., Apophis \citep{Pravec2014,Yu2017,brozovic2018}. Although the final spin state of this case does not match the actual state of Toutatis, this does not weaken our conclusion, because the asteroid's rotation may be altered by other mechanisms, such as a continual changing by the YORP effect or an instantaneous high-speed impact.

The outcomes show that collapse of materials from the head to the body is very common during the impact process. This is possibly because cohesion is not included in our model and explains why the sharp angle near the contact area is not observed in many cases. Recent research indicates that even a very small cohesive force can dramatically change the allowable spin rates for an asteroid \citep{sanchez2014strength}. The effect of cohesion in a low-speed impact scenario will be explored in our future studies.

In brief, our results given in this paper provide a likely formation scenario for 4179 Toutatis. However, this problem is still open since we cannot exclude other possible mechanisms, such as orbit contraction due to the secular BYORP effect in a synchronous binary system, or low-speed impact resulting from a large-scale catastrophic collision \citep{durda2004formation}. Even the speculation that the elongation is primordial (such as an elongated fragment ejected right after a large-scale catastrophic collision) is not excluded. In view of the still-puzzling slow non-principal rotation of Toutatis with the rotation axis along the long axis, further detailed investigations combined with cohesion and other possible mechanisms are warranted. Finally, bear in mind that the granular paradigm for the dynamical behavior of granular asteroid is not yet well established. Laying a solid foundation for the theoretical framework is necessary for more reliable numerical modeling.

\section*{Acknowledgements}
We appreciate Professor Daniel Scheeres for his insightful suggestions and comments. This work is financially supported by the National Natural Science Foundation of China (Grant Nos. 11473073, 11503091, 11661161013, 11633009, 11673072), CAS Interdisciplinary Innovation Team, and Foundation of Minor Planets of the Purple Mountain Observatory.

%%%%%%%%%%%%%%%%%%%%%%%%%%%%%%%%%%%%%%%%%%%%%%%%%%

%%%%%%%%%%%%%%%%%%%% REFERENCES %%%%%%%%%%%%%%%%%%

% Don't change these lines
\bsp	% typesetting comment
\label{lastpage}

\begin{thebibliography}{}

\bibitem[\protect\citeauthoryear{Barucci et al.}{2015}]{Barucci2015}
Barucci M.~A., Fulchignoni M., Ji J., Marchin S., Thomas N., 2015, Asteroids IV, Patrick Michel, Francesca E. DeMeo, and William F. Bottke (eds.), University of Arizona Press, Tucson, pp.433-450

\bibitem[\protect\citeauthoryear{Benner, Nolan, Ostro, Giorgini, Pray, Harris,
  Magri  \& Margot}{Benner et~al.}{2006}]{benner2006near}
Benner L.~A.,  Nolan M.~C.,  Ostro S.~J.,  Giorgini J.~D.,  Pray D.~P.,  Harris
  A.~W.,  Magri C.,   Margot J.-L.,  2006, Icarus, 182, 474

\bibitem[\protect\citeauthoryear{Benner et al.}{2015}]{Benner2015}
Benner L.~A.~M., Busch M.~W., Giorgini J.~D., Taylor P.~A., Margot J.-L., 2015, Asteroids IV, Patrick Michel, Francesca E. DeMeo, and William F. Bottke (eds.), University of Arizona Press, Tucson, pp.165-182

\bibitem[\protect\citeauthoryear{Bottke, Morbidelli, Jedicke, Petit, Levison,
  Michel  \& Metcalfe}{Bottke et~al.}{2002}]{bottke2002debiased}
Bottke W.~F.,  Morbidelli A.,  Jedicke R.,  Petit J.-M.,  Levison H.~F.,
  Michel P.,   Metcalfe T.~S.,  2002, Icarus, 156, 399

\bibitem[\protect\citeauthoryear{Bottke, Vokrouhlick{\`y}, Rubincam  \&
  Nesvorn{\`y}}{Bottke et~al.}{2006}]{bottke2006yarkovsky}
Bottke W.~F.,  Vokrouhlick{\`y} D.,  Rubincam D.~P.,   Nesvorn{\`y} D.,  2006,
  Annu. Rev. Earth Planet. Sci., 34, 157

\bibitem[\protect\citeauthoryear{Brozovi{\'c} et~al.,}{Brozovic
  et~al.}{2010}]{brozovic2010radar}
Brozovi{\'c} M.,  et~al., 2010, Icarus, 208, 207

\bibitem[\protect\citeauthoryear{Brozovi{\'c} et al.}{2018}]{brozovic2018}
Brozovi{\'c} M., et al., 2018, Icarus, 300, 115

\bibitem[\protect\citeauthoryear{Bu et~al.,}{Bu et~al.}{2014}]{bu2014new}
Bu Y.,  et~al., 2014, The Astronomical Journal, 149, 21

\bibitem[\protect\citeauthoryear{Busch et~al.,}{Busch
  et~al.}{2012}]{busch2012radar}
Busch M.~W.,  et~al., 2012, in AAS/Division for Planetary Sciences Meeting Abstracts.
\bibitem[\protect\citeauthoryear{Chau, Wong  \& Wu}{Chau
  et~al.}{2002}]{Chau2002coefficient}
Chau K.~T.,  Wong R.~H.~C., Wu J., 2002, International Journal of Rock Mechanics and Mining Sciences, 39, 69

\bibitem[\protect\citeauthoryear{Chauvineau, Mignard  \& Farinella}{Chauvineau
  et~al.}{1991}]{chauvineau1991lifetime}
Chauvineau B.,  Mignard F.,  Farinella P.,  1991, Icarus, 94, 299

\bibitem[\protect\citeauthoryear{Chauvineua, Farinella  \& Harris}{Chauvineua
  et~al.}{1995}]{chauvineua1995evolution}
Chauvineua B.,  Farinella P.,   Harris A.,  1995, Icarus, 115, 36

\bibitem[\protect\citeauthoryear{Da, Emam, Prochnow, Roux \& Chevoir}
{Da et~al.}{2005}]{Da2005rheophysics}
Da C.~F., Emam S., Prochnow M., Roux JN., 2005, Physical Review E, 72, 021309

\bibitem[\protect\citeauthoryear{Durda, Bottke, Enke, Merline, Asphaug,
  Richardson  \& Leinhardt}{Durda et~al.}{2004}]{durda2004formation}
Durda D.~D.,  Bottke W.~F.,  Enke B.~L.,  Merline W.~J.,  Asphaug E.,
  Richardson D.~C.,   Leinhardt Z.~M.,  2004, Icarus, 170, 243

\bibitem[\protect\citeauthoryear{Estrada, Az¨¦ma, Radjai \& Taboada}
{Estrada et~al.}{2011}]{Estrada2011identification}
Estrada N., Az¨¦ma E., Radjai F., Taboada A., 2011, Physical Review E, 84, 011306

\bibitem[\protect\citeauthoryear{Fang \& Margot}{Fang \&
  Margot}{2011}]{fang2011binary}
Fang J.,  Margot J.-L.,  2011, The Astronomical Journal, 143, 25

\bibitem[\protect\citeauthoryear{Fang, Margot, Brozovic, Nolan, Benner  \&
  Taylor}{Fang et~al.}{2011}]{fang2011orbits}
Fang J.,  Margot J.-L.,  Brozovic M.,  Nolan M.~C.,  Benner L.~A.,   Taylor
  P.~A.,  2011, The Astronomical Journal, 141, 154

\bibitem[\protect\citeauthoryear{Farinella}{Farinella}{1992}]{farinella1992evolution}
Farinella P.,  1992, Icarus, 96, 284

\bibitem[\protect\citeauthoryear{Farinella \& Chauvineau}{Farinella \&
  Chauvineau}{1993}]{farinella1993evolution}
Farinella P.,  Chauvineau B.,  1993,  Astronomy and Astrophysics, 279, 251

\bibitem[\protect\citeauthoryear{Gladman, Michel  \& Froeschl{\'e}}{Gladman
  et~al.}{2000}]{gladman2000near}
Gladman B.,  Michel P.,   Froeschl{\'e} C.,  2000, Icarus, 146, 176

\bibitem[\protect\citeauthoryear{Harmon, Nolan, Giorgini  \& Howell}{Harmon
  et~al.}{2010}]{harmon2010radar}
Harmon J.~K.,  Nolan M.~C.,  Giorgini J.~D.,   Howell E.~S.,  2010, Icarus,
  207, 499

\bibitem[\protect\citeauthoryear{Holsapple \& Housen}{Holsapple \&
  Housen}{2007}]{holsapple2007crater}
Holsapple K.~A.,  Housen K.~R.,  2007, Icarus, 191, 586

\bibitem[\protect\citeauthoryear{Hu \& Ji}{2017}]{Hu2017}
Hu S. C., Ji J., 2017, RAA, 17, 120

\bibitem[\protect\citeauthoryear{Huang et~al.,}{Huang
  et~al.}{2013a}]{huang2013ginger}
Huang J.,  et~al., 2013a, Scientific Reports, 3, 3411

\bibitem[\protect\citeauthoryear{Huang et~al.}{Huang
  et~al.}{2013b}]{huang2013engineering}
Huang J.,  et~al., 2013b, Science China Technological Sciences, 43, 596

\bibitem[\protect\citeauthoryear{Hudson \& Ostro}{Hudson \&
  Ostro}{1995}]{hudson1995shape}
Hudson R.~S.,  Ostro S.~J.,  1995, Science, 270, 84

\bibitem[\protect\citeauthoryear{Hudson \& Ostro}{Hudson \&
  Ostro}{1998}]{hudson1998photometric}
Hudson R.~S.,  Ostro S.~J.,  1998, Icarus, 135, 451

\bibitem[\protect\citeauthoryear{Hudson, Ostro  \& Scheeres}{Hudson
  et~al.}{2003}]{hudson2003high}
Hudson R.,  Ostro S.,   Scheeres D.,  2003, Icarus, 161, 346

\bibitem[\protect\citeauthoryear{Ji \& Liu}{2001}]{Ji2001}
Ji J., Liu L., 2001, ChA\&A, 25, 147

\bibitem[\protect\citeauthoryear{Ji et al.}{2016}]{Ji2016}
Ji J., Jiang Y., Zhao Y., Wang S., Yu L., 2016, IAU Symposium, 318, 144

\bibitem[\protect\citeauthoryear{Jiang, Ji, Huang, Marchi, Li  \& Ip}{Jiang
  et~al.}{2015}]{jiang2015boulders}
Jiang Y.,  Ji J.,  Huang J.,  Marchi S.,  Li Y.,  Ip W.-H.,  2015, Scientific
  Reports, 5,  16029

\bibitem[\protect\citeauthoryear{Jiang, Shen  \& Wang}{Jiang
  et~al.}{2015}]{jiang2015novel}
Jiang M.,  Shen Z.,   Wang J.,  2015, Computers and Geotechnics, 65, 147


\bibitem[\protect\citeauthoryear{Jutzi \& Asphaug}{Jutzi \&
  Asphaug}{2015}]{jutzi2015shape}
Jutzi M.,  Asphaug E.,  2015, Science, 348, 1355

\bibitem[\protect\citeauthoryear{Jutzi \& Benz}{Jutzi \&
  Benz}{2017}]{jutzi2017formation}
Jutzi M.,  Benz W.,  2017, Astronomy \& Astrophysics, 597, A62

\bibitem[\protect\citeauthoryear{Krivova, Yagudina  \& Shor}{Krivova
  et~al.}{1994}]{krivova1994orbit}
Krivova N.,  Yagudina E.,   Shor V.,  1994, Planetary and Space Science, 42,
  741

\bibitem[\protect\citeauthoryear{Magri et~al.,}{Magri
  et~al.}{2011}]{magri2011radar}
Magri C.,  et~al., 2011, Icarus, 214, 210

\bibitem[\protect\citeauthoryear{Michel, Froeschl{\'e}  \& Farinella}{Michel
  et~al.}{1996}]{michel1996dynamical}
Michel P.,  Froeschl{\'e} C.,   Farinella P.,  1996, Earth, Moon, and Planets,
  72, 151

\bibitem[\protect\citeauthoryear{Morbidelli \& Vokrouhlick{\`y}}{Morbidelli \&
  Vokrouhlick{\`y}}{2003}]{morbidelli2003yarkovsky}
Morbidelli A.,  Vokrouhlick{\`y} D.,  2003, Icarus, 163, 120

\bibitem[\protect\citeauthoryear{Ostro et~al.,}{Ostro
  et~al.}{1995}]{ostro1995radar}
Ostro S.~J.,  et~al., 1995, Science, 270, 80

\bibitem[\protect\citeauthoryear{Ostro et~al.,}{Ostro
  et~al.}{1999}]{ostro1999asteroid}
Ostro S.~J.,  et~al., 1999, Icarus, 137, 122

\bibitem[\protect\citeauthoryear{Pravec \& Harris}{Pravec \&
  Harris}{2007}]{pravec2007binary}
Pravec P.,  Harris A.~W.,  2007, Icarus, 190, 250

\bibitem[\protect\citeauthoryear{Pravec et al.}{2014}]{Pravec2014}
Pravec P., et al., 2014, Icarus, 233, 48


\bibitem[\protect\citeauthoryear{Reddy, Sanchez, Gaffey, Abell, Le~Corre  \&
  Hardersen}{Reddy et~al.}{2012}]{reddy2012composition}
Reddy V.,  Sanchez J.~A.,  Gaffey M.~J.,  Abell P.~A.,  Le~Corre L.,
  Hardersen P.~S.,  2012, Icarus, 221, 1177

\bibitem[\protect\citeauthoryear{Richardson, Bottke  \& Love}{Richardson
  et~al.}{1998}]{richardson1998tidal}
Richardson D.~C.,  Bottke W.~F.,   Love S.~G.,  1998, Icarus, 134, 47

\bibitem[\protect\citeauthoryear{Richardson, Quinn, Stadel  \& Lake}{Richardson
  et~al.}{2000}]{richardson2000direct}
Richardson D.~C.,  Quinn T.,  Stadel J.,   Lake G.,  2000, Icarus, 143, 45

\bibitem[\protect\citeauthoryear{Richardson, Michel, Walsh  \&
  Flynn}{Richardson et~al.}{2009}]{richardson2009numerical}
Richardson D.~C.,  Michel P.,  Walsh K.~J.,   Flynn K.,  2009, Planetary and
  Space Science, 57, 183

\bibitem[\protect\citeauthoryear{Richardson, Walsh, Murdoch  \&
  Michel}{Richardson et~al.}{2011}]{richardson2011numerical}
Richardson D.~C.,  Walsh K.~J.,  Murdoch N.,   Michel P.,  2011, Icarus, 212,
  427

\bibitem[\protect\citeauthoryear{S{\'a}nchez \& Scheeres}{S{\'a}nchez \&
  Scheeres}{2014}]{sanchez2014strength}
S{\'a}nchez P.,  Scheeres D.~J.,  2014, Meteoritics \& Planetary Science, 49,
  788

\bibitem[\protect\citeauthoryear{Scheeres}{Scheeres}{2007}]{scheeres2007rotational}
Scheeres D.~J.,  2007, Icarus, 189, 370

\bibitem[\protect\citeauthoryear{Scheeres, Ostro, Werner, Asphaug  \&
  Hudson}{Scheeres et~al.}{2000}]{scheeres2000effects}
Scheeres D.~J.,  Ostro S.~J.,  Werner R.~A.,  Asphaug E.,   Hudson R.~S.,
  2000, Icarus, 147, 106

\bibitem[\protect\citeauthoryear{Schwartz, Richardson  \& Michel}{Schwartz
  et~al.}{2012}]{schwartz2012implementation}
Schwartz S.~R.,  Richardson D.~C.,  Michel P.,  2012, Granular Matter, 14, 363

\bibitem[\protect\citeauthoryear{Schwartz, Michel, Jutzi, Marchi \& Zhang}{Schwartz
 et~al.}{2018}]{schwartz2018catastrophic}
Schwartz S.~R.,  Michel P.,  Jutzi M.,  Marchi S.,  Zhang Y.,  Richardson D.~C.,  2018,  Nature Astronomy, DOI:10.1038/s41550-018-0395-2


\bibitem[\protect\citeauthoryear{Sitarski}{Sitarski}{1998}]{sitarski1998motion}
Sitarski G.,  1998, Acta Astronomica, 48, 547

\bibitem[\protect\citeauthoryear{Spencer et~al.,}{Spencer
  et~al.}{1995}]{spencer1995lightcurve}
Spencer J.~R.,  et~al., 1995, Icarus, 117, 71

\bibitem[\protect\citeauthoryear{Stadel}{Stadel}{2001}]{stadel2001cosmological}
Stadel J.~G.,  2001, PhD thesis, University of Washington Washington, DC

\bibitem[\protect\citeauthoryear{Takahashi, Busch  \& Scheeres}{Takahashi
  et~al.}{2013}]{takahashi2013spin}
Takahashi Y.,  Busch M.~W.,   Scheeres D.,  2013, The Astronomical Journal,
  146, 95

\bibitem[\protect\citeauthoryear{Walsh \& Jacobson}{Walsh \&
  Jacobson}{2015}]{walsh2015formation}
Walsh K.~J.,  Jacobson S.~A.,  2015, Asteroids IV, Patrick Michel,
Francesca E. DeMeo, and William F. Bottke (eds.), University of Arizona Press, Tucson, pp.375--393

\bibitem[\protect\citeauthoryear{Walsh \& Richardson}{Walsh \&
  Richardson}{2006}]{walsh2006binary}
Walsh K.~J.,  Richardson D.~C.,  2006, Icarus, 180, 201

\bibitem[\protect\citeauthoryear{Walsh \& Richardson}{Walsh \&
  Richardson}{2008}]{walsh2008steady}
Walsh K.~J.,  Richardson D.~C.,  2008, Icarus, 193, 553

\bibitem[\protect\citeauthoryear{Whipple \& Shelus}{Whipple \&
  Shelus}{1993}]{whipple1993long}
Whipple A.~L.,  Shelus P.~J.,  1993, Icarus, 105, 408

\bibitem[\protect\citeauthoryear{Yu, Ji, \& Ip}{2017}]{Yu2017}
Yu L.-L., Ji J., Ip W.-H., 2017, RAA, 17, 070

\bibitem[\protect\citeauthoryear{Yu, Richardson, Michel, Schwartz  \&
  Ballouz}{Yu et~al.}{2014}]{yu2014numerical}
Yu Y.,  Richardson D.~C.,  Michel P.,  Schwartz S.~R.,   Ballouz R.-L.,  2014,
  Icarus, 242, 82

\bibitem[\protect\citeauthoryear{Zhang et~al.,}{Zhang et~al.}{2017}]{zhang2017creep}
Zhang Y.,  et~al., 2017, Icarus, 294, 98

\bibitem[\protect\citeauthoryear{Zhang, Richardson, Barnouin, Michel, Schwartz \& Ballouz}
{Zhang et~al.}{2018}]{zhang2018rotational}
Zhang Y., Richardson D.~C.,  Barnouin S.~O., Michel P., Schwartz S.~R., Ballouz R.-L., 2018, The
Astrophysical Journal,  857, 15

\bibitem[\protect\citeauthoryear{Zhao, Ji, Huang, Hu, Hou, Li  \& Ip}{Zhao
  et~al.}{2015}]{zhao2015orientation}
Zhao Y.,  Ji J.,  Huang J.,  Hu S.,  Hou X.,  Li Y.,   Ip W.-H.,  2015, Monthly
  Notices of the Royal Astronomical Society, 450, 3620

\bibitem[\protect\citeauthoryear{Zhao, Xiao, Liu, Sun, Huang  \& Tang}{Zhao
  et~al.}{2016}]{zhao2016radar}
Zhao W.,  Xiao T.,  Liu P.,  Sun L.,  Huang J.,   Tang X.,  2016, Planetary and
  Space Science, 125, 87

\bibitem[\protect\citeauthoryear{Zhu et~al.,}{Zhu
  et~al.}{2014}]{zhu2014morphology}
Zhu M.-H.,  et~al., 2014, Geophysical Research Letters, 41, 328

\bibitem[\protect\citeauthoryear{Zou, Li, Liu, Wang, Li  \& Ping}{Zou
  et~al.}{2014}]{zou2014preliminary}
Zou X.,  Li C.,  Liu J.,  Wang W.,  Li H.,   Ping J.,  2014, Icarus, 229, 348

\makeatother
\end{thebibliography}
\end{document}